\documentclass[aps,prd,twocolumn,superscriptaddress,nofootinbib,floatfix]{revtex4-1}

\usepackage{latexsym}
\usepackage{amsmath}
\usepackage{amssymb}
\usepackage{amsfonts}
\usepackage{bm}
\usepackage{physics}
\usepackage{slashed}

\usepackage{color}
\definecolor{purple}{rgb}{0.5,0,0.5}
\definecolor{blue}{rgb}{0.0,0,0.9}
\definecolor{prdblue}{rgb}{0.133,0.118,0.498}
\usepackage[colorlinks=true, pdfstartview=FitV, linkcolor=prdblue, citecolor= prdblue, urlcolor=prdblue]{hyperref}

\usepackage{supertabular} 
\usepackage{placeins}
\usepackage{epsfig}
\usepackage{graphicx}

\begin{document}


\title{Leading-twist proton parton distribution amplitude \\ from scalar and axial-vector diquark correlations}


\author{B. Almeida-Zamora}
\email[]{bilgai\_almeidaz@hotmail.com}
\affiliation{Departamento de Sistemas F\'isicos, Qu\'imicos y Naturales, Universidad Pablo de Olavide, Avenida Rectora Rosario Valpuesta 1, 41089, Dos Hermanas, Sevilla, Spain.}
\affiliation{Departamento de F\'isica, Universidad de Sonora, Boulevard Luis Encinas J. y Rosales, 83000, Hermosillo, Sonora, Mexico.}

\author{L. Albino}
\email[]{luis.albino.fernandez@gmail.com}
\affiliation{Departamento de F\'isica, Universidad de Sonora, Boulevard Luis Encinas J. y Rosales, 83000, Hermosillo, Sonora, Mexico.}

\author{C. Mezrag}
\email[]{cedric.mezrag@cea.fr}
\affiliation{IRFU, CEA, Université Paris-Saclay, 91191 Gif-sur-Yvette, France}
\affiliation{AIDAS, CEA, 91191, Gif-Sur-Yvette, France}

\author{J. Segovia}
\email[]{jsegovia@upo.es}
\affiliation{Departamento de Sistemas F\'isicos, Qu\'imicos y Naturales, Universidad Pablo de Olavide, Avenida Rectora Rosario Valpuesta 1, 41089, Dos Hermanas, Sevilla, Spain.}


\date{\today}

\begin{abstract}
We compute the leading-twist proton parton distribution amplitude (PDA) using algebraic representations of the nucleon Faddeev wave function inspired by perturbation-theory integral representations. The PDA is obtained from a light-front projection of the covariant quark--diquark Faddeev amplitude and analysed in terms of its scalar- and axial-vector-diquark components. We derive analytic expressions for the corresponding diquark distribution amplitudes and their contributions to the proton PDA. The resulting distribution is broad and concave, with its maximum $(x_1,x_2,x_3)\approx(0.55,0.20,0.25)$ displaced from the symmetric configuration $(1/3,1/3,1/3)$, which corresponds to the maximum of the perturbative-QCD asymptotic form. Scalar-diquark correlations generate a strong asymmetry in the longitudinal momentum distribution, whereas axial-vector correlations redistribute momentum among the valence quarks and significantly reduce this asymmetry. We calculate low-order Mellin moments and obtain a compact parametrisation of the proton PDA at the hadronic scale, $\mu_0=0.51\,\mathrm{GeV}$. We then perform the leading-order QCD evolution of its normalized shape to $\mu=2\,\mathrm{GeV}$, finding that evolution reduces the momentum asymmetry while preserving a clear imprint of the non-perturbative quark--diquark correlations. The results provide insight into the role of such correlations in shaping the proton light-front structure and furnish non-perturbative input for studies of hard exclusive baryonic processes.
\end{abstract}


\maketitle


\section{Introduction}
\label{sec:intro}

Understanding how the structure and properties of hadrons emerge from the dynamics of quarks and gluons in quantum chromodynamics (QCD) remains a central challenge in strong-interaction physics. In hard exclusive processes, factorisation theorems separate short-distance dynamics, calculable in perturbative QCD, from universal non-perturbative quantities that encode hadron structure. Among these, parton distribution amplitudes (PDAs) play a key role because they determine the longitudinal momentum distribution of valence partons in the leading Fock component of a hadron and provide essential input for the description of exclusive reactions at large momentum transfer~\cite{Efremov:1979qk, Lepage:1980fj, Chernyak:1984bm}.

The asymptotic behaviour of PDAs at very large momentum scales has been known for several decades. However, determining their shape at experimentally accessible energies remains a non-trivial problem. In the meson sector, continuum and lattice-QCD studies have established that the PDAs of light pseudoscalar mesons are broad, concave distributions that differ substantially from their asymptotic forms at experimentally accessible scales~\cite{Chang:2013pq, Gao:2014bca, Shi:2015esa, Segovia:2013eca, Braun:2015axa, Bali:2017ude, Zhang:2017bzy}.

In contrast, considerably less is known about baryon PDAs. Several theoretical methods have been employed to estimate the proton PDA, including QCD sum rules~\cite{Chernyak:1983ej, Stefanis:1992nw}, static diquark models~\cite{Dziembowski:1990md}, phenomenological parametrisations~\cite{Bolz:1996sw}, light-front quark models~\cite{Pasquini:2009ki}, light-cone sum rules~\cite{Anikin:2013aka}, and lattice-QCD calculations that have provided increasingly precise determinations of low Mellin moments of the proton and octet-baryon PDAs~\cite{RQCD:2019hps,Bali:2024oxg}. More recently, lattice-QCD approaches based on large-momentum effective theory have begun to provide access to the full $x$ dependence of baryon distribution amplitudes~\cite{LPC:2026mvw}. These studies broadly support a distribution that differs substantially from the asymptotic form; however, the dynamical origin of its momentum asymmetry remains incompletely understood. In particular, it is important to determine how the non-perturbative quark--quark correlations that emerge in continuum studies of baryons influence the shape of the proton PDA.

Over the past decades, continuum approaches to QCD have provided important new insights into baryon structure. In particular, studies based on Dyson--Schwinger equations and covariant Faddeev equations have demonstrated that strong diquark correlations are dynamically generated inside baryons~\cite{Cahill:1988dx, Burden:1988dt, Cahill:1988zi, Reinhardt:1989rw, Efimov:1990uz}. Within this framework, the nucleon can be understood as a Borromean bound state of three dressed quarks in which dynamically generated scalar and axial-vector diquark correlations play a dominant role~\cite{Segovia:2015ufa}. This description is supported by extensive studies of nucleon elastic and transition form factors, which have successfully reproduced a wide range of experimental data \cite{Segovia:2013uga, Segovia:2013rca, Segovia:2014aza, Segovia:2015hra, Segovia:2016zyc}.

A complementary development has been the use of perturbation-theory integral representations (PTIRs), originally introduced by Nakanishi~\cite{Nakanishi:1963zz, Nakanishi:1969ph}, as a framework for constructing analytic representations of Bethe--Salpeter and Faddeev amplitudes. Within continuum QCD, PTIR-based approaches have enabled analytic studies of a variety of hadron structure observables, including parton distribution functions, generalized parton distributions, and distribution amplitudes~\cite{Chang:2014lva, Mezrag:2014jka, Mezrag:2014tva, Mezrag:2016hnp, Chouika:2017dhe, Chouika:2017rzs}. In the baryon sector, this framework has recently been used to obtain a continuum-QCD calculation of the proton PDA~\cite{Mezrag:2017znp, Mezrag:2018hkk}. However, the separate roles of scalar- and axial-vector-diquark correlations in shaping the resulting distribution have not yet been analysed in detail.

While Refs.~\cite{Mezrag:2017znp, Mezrag:2018hkk} provided a PTIR-based calculation of the proton PDA, the separate roles of scalar- and axial-vector-diquark correlations were not analysed in detail. The present work addresses this question within an algebraic PTIR-inspired framework that permits an explicit decomposition of the proton PDA into its underlying quark--diquark components. This makes it possible to identify how different diquark channels contribute to the longitudinal momentum distribution of the valence quarks and to quantify their impact on the resulting Mellin moments and PDA shape.

Determining the shape of the proton PDA is important for improving theoretical descriptions of hard exclusive processes involving baryons. Since the PDA encodes the longitudinal momentum distribution of valence quarks, it provides a direct link between non-perturbative descriptions of nucleon structure and perturbative calculations of scattering amplitudes. The PTIR-inspired algebraic framework employed in this work enables an analytic treatment of the underlying quark--diquark dynamics, making it possible to investigate how scalar- and axial-vector-diquark correlations shape the proton PDA and redistribute longitudinal momentum among the valence quarks.

This manuscript is organised as follows. In Sec.~\ref{sec:theory} we introduce the theoretical framework, including the definition of baryon distribution amplitudes, the quark--diquark Faddeev wave function, and the connection between Mellin moments and light-front projections. In Sec.~\ref{sec:scalardiquarkonly} we analyse the scalar-diquark contribution and construct the corresponding proton PDA. Section~\ref{sec:axialvector} extends the calculation to include axial-vector diquark correlations and presents the corresponding proton PDA. The complete PDA of the proton and its associated Mellin moments are discussed in Sec.~\ref{sec:PDAtotal}. Finally, Sec.~\ref{sec:summary} summarises our conclusions and outlines possible future developments.


\section{Formalism}
\label{sec:theory}

\subsection{Parton distribution amplitude}
\label{subsec:PDAdefinition}

Hadron distribution amplitudes are defined through vacuum-to-hadron matrix elements of non-local operators composed of quark and gluon fields separated along light-like directions~\cite{Chernyak:1984bm, Braun:2000kw}. In the present work we consider a three--quark operator evaluated along a light-like vector $n$, which allows us to define the leading-twist PDA of the nucleon. Throughout this manuscript the hadron momentum is denoted by $P$, while its polarisation state is labelled by $\lambda$. Besides, it is convenient to introduce an auxiliary light-like vector $p$ defined as
\begin{subequations}
\begin{eqnarray}
\label{eq:LightConeVector}
p^\mu & = & P^\mu - n^\mu \frac{M_N^2}{2 P\cdot n}, \\
p \cdot n & = & P \cdot n\ ,
\end{eqnarray}
\end{subequations}
where $M_N$ denotes the nucleon mass.

Using the introduced vectors above, one may express the matrix element of the three--quark operator in terms of a set of invariant functions that correspond to different Dirac structures. Omitting Wilson lines for simplicity and denoting the transposed quark field by $\tilde{u}$, the leading decomposition reads~\cite{Braun:2000kw, Diehl:2003ny}
\begin{align}
\label{eq:LeadingTwistContributions}
\bra{0}  \epsilon^{ijk} & \tilde{u}_\alpha^{i}(z_1 n) u_\beta^{j}(z_2 n) d_\gamma^{k}(z_3 n)\ket{P,\lambda} = \nonumber\\
&
\frac{f_N}{4} \left[ \left(\slashed{p} C \right)_{\alpha \beta}\left( \gamma_5 N^+ \right)_{\gamma} V(z_i  p\cdot n)  \nonumber \right.\\
&
\left. + \left(\slashed{p}\gamma_5 C \right)_{\alpha\beta}\left( N^+ \right)_\gamma A(z_i p\cdot n) \right. \nonumber \\ 
&
\left.- \left(ip^\mu \sigma_{\mu\nu} C \right)_{\alpha\beta}\left( \gamma^\nu\gamma_5 N^+\right)_\gamma T(z_i p\cdot n) \right] \nonumber \\
&
+ \textrm{higher twist} \,.
\end{align}
That is to say, among the $24$ possible tensor structures appearing in the general decomposition, only three contribute at leading twist.\footnote{A full twist decomposition is available in Ref.~\cite{Braun:2000kw}.} In the expression above, $C$ denotes the charge-conjugation matrix and $N^+$ the so-called \emph{good} component of the nucleon spinor, \emph{i.e.} $N^+(P)$ fulfills $\slashed{p} N(P) = M N^+(P)$, where $N(P)$ is the nucleon spinor. The constant $f_N$ corresponds to the value of the nucleon wave function at the origin. Greek indices $(\alpha,\beta,\gamma)$ refer to Dirac components whereas latin indices $(i,j,k)$ label colour degrees of freedom. The scalar functions $V$, $A$, and $T$ depend exclusively on the light-cone coordinates $(z_1,z_2,z_3)$ along the direction $n$. To express these functions in terms of momentum fractions it is convenient to introduce their Fourier transforms:
\begin{equation}
\label{eq:FourierTransform}
F(z_i p \cdot n ) = \int \mathcal{D}x \, e^{-i p\cdot n \sum_{i} x_i z_i} F(x_i),
\end{equation}
where $\mathcal{D}x = \left(\prod_i\textrm{d}x_i\right) \delta(1-\sum_j x_j)$.

The leading--twist proton PDA is then defined through the combination
\begin{equation}
\label{eq:PhiDef}
\varphi(\left[x\right]) = V([x]) - A([x]).
\end{equation}
with $[x]\equiv(x_1,x_2,x_3)$.

This definition is particularly useful because the functions $V$, $A$, and $T$ are not independent. Owing to the presence of two identical $u$ quarks in the proton, symmetry under their interchange constrains the functional form of these amplitudes. In particular, $V$ is symmetric while $A$ is antisymmetric under the exchange $x_1 \leftrightarrow x_2$, and thus
\begin{subequations}
\begin{align}
\label{eq:Symmetries}
\varphi(x_1,x_2,x_3) = & V(x_1,x_2,x_3) - A(x_1,x_2,x_3), \\
\varphi(x_2,x_1,x_3) = & V(x_1,x_2,x_3) + A(x_1,x_2,x_3),\\
2 T(x_1,x_2,x_3) = &   \varphi(x_1,x_3,x_2)+  \varphi(x_2,x_3,x_1),
\end{align}    
\end{subequations}
where the last relation follows when isospin symmetry is assumed. These relations show that the essential dynamical information of the leading-twist proton PDA is fully encoded in the function $\varphi(\left[x\right])$.

The physical interpretation of the PDA becomes transparent when the proton state is expanded along the light front. At leading twist the Fock decomposition of the proton state can be written as
\begin{align}
\label{eq:ProtonLT}
\ket{P,\uparrow} &= \int \frac{{\cal D}x}{8\sqrt{6x_1x_2x_3}}\ket{uud}\otimes \left[\varphi(x_1,x_2,x_3)\ket{\uparrow \downarrow \uparrow} \right. \nonumber\\
&
\left. + \varphi(x_2,x_1,x_3)\ket{\downarrow \uparrow \uparrow} \!-\!2T(x_1,x_2,x_3)\ket{\uparrow \uparrow \downarrow} \right],
\end{align} 
This equation makes explicit that the leading--twist PDA characterises the distribution of longitudinal momentum among the valence quarks. In this approximation orbital angular momentum contributions are absent.

For practical calculations it is useful to express the PDA through matrix elements of suitable projected operators. A convenient projection of Eq.~\eqref{eq:LeadingTwistContributions} yields
\begin{align}
\label{eq:PhiasMatrix1}
\bra{0}&\epsilon^{ijk}\left(\tilde{u}_\uparrow^{i}(z_1)C\slashed{n}u_\downarrow^{j}(z_2)\right)\slashed{n}d_\uparrow^{k}(z_3)\ket{P,\lambda} = \\
&-\frac{1}{2} (p \cdot n)f_N \slashed{n} N^\uparrow\int \mathcal{D}x \varphi(x_1,x_2,x_3) e^{-ip \cdot n\sum_i x_i z_i},\nonumber
\end{align}
where $q^{\uparrow (\downarrow)} = \frac{1 \pm \gamma_5}{2}q$ denotes a quark field projected onto definite chirality and $N^\uparrow$ is the corresponding component of the nucleon spinor.

For completeness we also display the matrix-element definition of the function $T$, which we compute independently as a consistency check of the isospin relations:
\begin{align}\label{eq:TasMatrix1}
    \bra{0}&\epsilon^{ijk}\left(u_\uparrow^{i}(z_1)Ci\sigma_{\perp \nu}n^\nu u_\uparrow^{j}(z_2)\right)\gamma^\perp\slashed{n}d_\downarrow^{k}(z_3)\ket{P,\lambda} = \\
    &-2 (p \cdot n)f_N \slashed{n} N^\uparrow\int \mathcal{D}x_i T(x_1,x_2,x_3) e^{-ip \cdot n\sum_i x_i z_i}, \nonumber
\end{align}
with $\gamma^\perp \gamma_\perp = \gamma^\mu\gamma^\nu \eta_{\mu\nu}^\perp$ and
\begin{equation}
  \label{eq:etaperp}
  \eta_{\mu\nu}^\perp = \eta_{\mu \nu}-\frac{n^\mu p^\nu + n^\nu p^\mu}{p\cdot n}.
\end{equation}

Finally, we recall that the operator in Eq.~\eqref{eq:PhiasMatrix1} requires renormalisation, which introduces a dependence on the renormalisation scale $\mu$. Consequently, the PDA $\varphi([x])$ should be understood as a scale-dependent quantity, $\varphi([x],\mu^2)$, whose evolution is governed by perturbative QCD evolution equations~\cite{Lepage:1980fj, Braun:2006hz}. For reference, the asymptotic form reached in the asymptotic $\mu^2 \to \infty$ is given by
\begin{equation}
\label{eq:AsymptoticPDA}
\varphi_{\mathrm{asy}}(x_1,x_2,x_3) \!=\! \lim_{\mu^2 \rightarrow \infty} \!\varphi(x_1,x_2,x_3,\mu^2) \!=\! 120\, x_1 x_2 x_3 \,.
\end{equation}


\begin{figure}[!t]
\centering
\includegraphics[width=\linewidth]{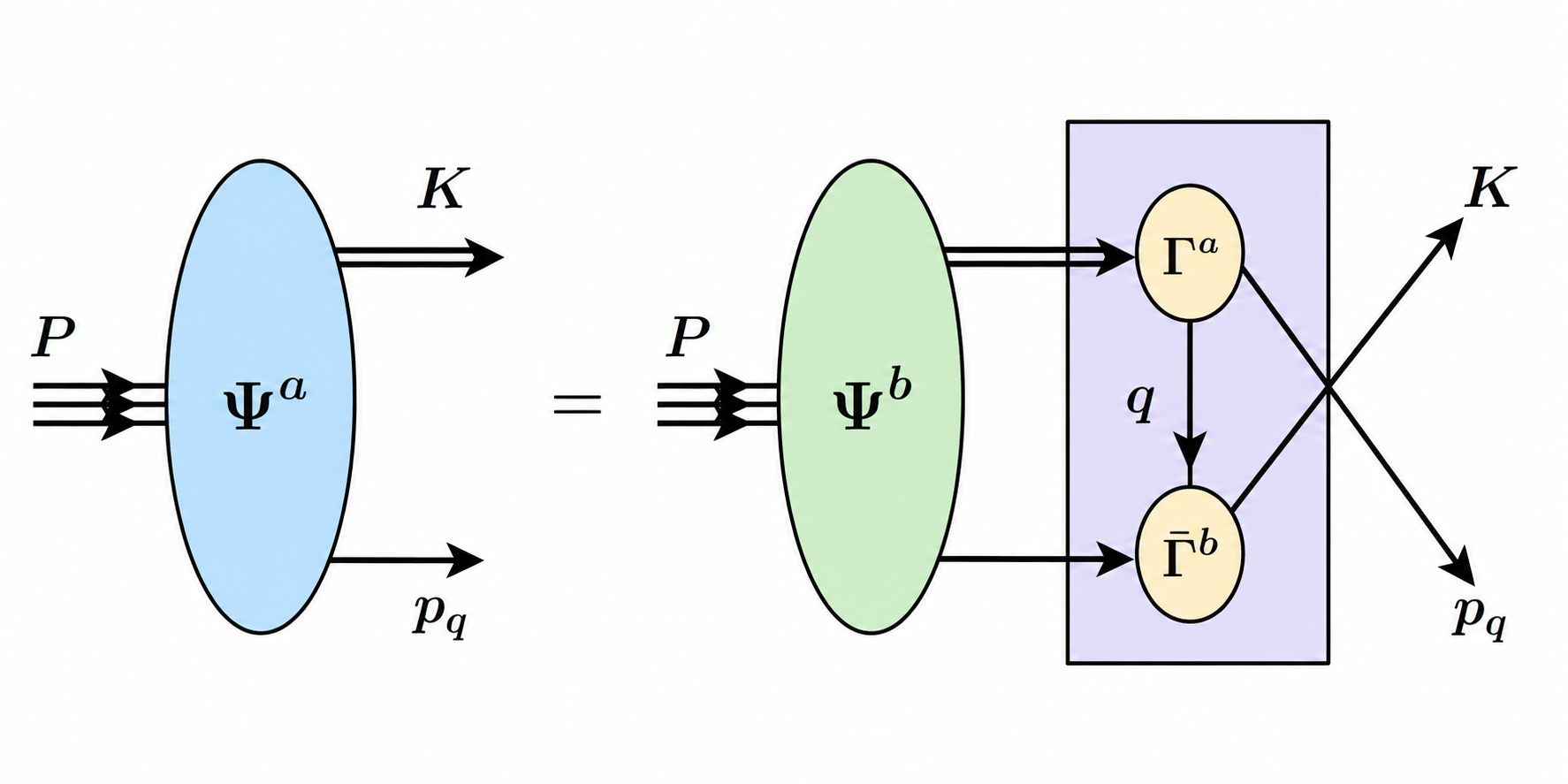}
\caption{\label{fig:FaddeevEquation} Poincaré-covariant Faddeev equation used in the present work to determine the properties of the nucleon. The baryon amplitude, $\Psi$, describes a bound state with total momentum $P=p_q+K$ and encodes the momentum correlations between the dressed quark and the diquark constituents. The purple box identifies the interaction kernel of the Faddeev equation. Within this kernel, single lines represent dressed-quark propagators, double lines denote diquark propagators, and $\Gamma$ corresponds to the diquark Bethe--Salpeter-like correlation amplitude.}
\end{figure}

\subsection{Faddeev wave function}
\label{subsec:Faddeev}

Continuum studies of the Faddeev equation reveal the emergence of strong, non-pointlike quark--quark correlations inside baryons, commonly referred to as diquarks. These dynamically generated correlations appear in attractive colour-$\bar 3$ channels and provide an efficient reorganisation of the three-body problem. Their presence is supported by extensive continuum analyses and lattice-QCD simulations~\cite{Alexandrou:2006cq, Babich:2007ah, Cloet:2013jya, Segovia:2015ufa, Barabanov:2020jvn}.

The dominant consequence of these correlations is that the three-valence-quark scattering problem can be reformulated as a Poincar\'e-covariant quark--diquark Faddeev equation, illustrated in Fig.~1. Notably, fully covariant three-body calculations that do not assume diquark clustering yield quantitatively similar results, indicating that diquarks emerge naturally as dominant correlations within baryons~\cite{Eichmann:2016yit}.

In the nucleon, the Faddeev amplitude is dominated by scalar ($0^+$) and axial-vector ($1^+$) diquark correlations. Consequently, the nucleon's rest-frame wave function is predominantly $S$-wave, with smaller contributions from higher partial waves~\cite{Cloet:2008re, Eichmann:2016yit}.

Taking into account the three possible choices for the bystander quark, the nucleon Faddeev amplitude $\Psi_N$ can be written as
\begin{equation}\label{eq:FaddeevDecomposition}
\Psi_N = \psi_1+\psi_2+\psi_3,
\end{equation}
where the subscript identifies the bystander quark, \textit{i.e.} the quark that does not participate in the diquark correlation. This decomposition ensures that all possible quark exchanges are included and that the full amplitude satisfies the required permutation symmetries.

Focusing on $\psi_3$, where quark~3 is the bystander and quarks~1 and 2 form the correlated pair, one can decompose
\begin{equation}
\label{eq:Psi3Decomposition}
\psi_3(p_i,\alpha_i,\tau_i) = \mathcal{N}^{0}_3 + \mathcal{N}^{1}_3,
\end{equation}
with $(p_i,\alpha_i,\tau_i)$ denoting the momenta, spinor and isospin indices of the constituent quarks. The total momentum is $P=p_1+p_2+p_3$, and the diquark momentum is $K=p_1+p_2$. The remaining components $\psi_1$ and $\psi_2$ are obtained by cyclic permutation.

The scalar-diquark contribution reads
\begin{equation}
\label{eq:FaddeevScalar}
\mathcal{N}^0_3 =
\left[\Gamma^0(k,K)\right]^{\alpha_1 \alpha_2}_{\tau_1 \tau_2}\,
\Delta^0(K)\,
\left[\mathcal{S}(\ell;P) N(P) \right]^{\alpha_3}_{\tau_3},
\end{equation}
where $\Gamma^0(k,K)$ is the Bethe--Salpeter amplitude of the scalar diquark, $\Delta^0(K)$ its propagator, and $\mathcal{S}(\ell;P)$ the scalar quark--diquark amplitude of the nucleon. Here, $k=p_1$ is a relative momentum within the diquark and $\ell=-K+2P/3$ is the quark--diquark relative momentum. This structure reflects the factorization of the Faddeev kernel into diquark formation, propagation, and subsequent coupling to the bystander quark.

The colour structure is antisymmetric, ensuring that the correlated quark pair lies in the attractive colour-$\bar 3$ channel. This can be expressed via the Levi-Civita tensor $\epsilon_{c_1 c_2 c_3}$, which may be written in terms of antisymmetric Gell-Mann matrices
\begin{subequations}
\begin{align}
\{H^1=i\lambda^7,\;H^2=-i\lambda^5,\;H^3&=i\lambda^2\}, \\
\epsilon_{c_1 c_2 c_3}= (H^{c_3})_{c_1 c_2}.
\end{align}    
\end{subequations}

The axial-vector contribution is given by
\begin{equation}
 \label{eq:FaddeevAV}
 \mathcal{N}^{1}_3 =
 \left[\Gamma^{j;1}_{\mu}(k,K)\right]^{\alpha_1 \alpha_2}_{\tau_1 \tau_2}
 \Delta^{1}_{\mu\nu}(K)
 \left[\mathcal{A}^j_\nu(\ell;P) N(P) \right]^{\alpha_3}_{\tau_3},
\end{equation}
where $\Gamma^{j;1}_{\mu}$ is the Bethe--Salpeter amplitude of an axial-vector diquark, with $\Delta^{1}_{\mu\nu}$ its propagator, and $\mathcal{A}^j_\nu(\ell;P)$ the corresponding nucleon's quark--diquark amplitude. The index $j$ labels the isospin-triplet nature of the axial-vector channel. Together, Eqs.~\eqref{eq:FaddeevScalar} and \eqref{eq:FaddeevAV} encode the dominant spin-flavour correlations of the nucleon and provide the covariant input required for the computation of its leading-twist PDA.

Finally, the Faddeev wave function $\chi_N$ is obtained from the amplitude $\Psi_N$ by attaching the dressed-quark and diquark propagators to the external legs. In this sense, $\Psi_N$ is the amputated bound-state amplitude, whereas $\chi_N$ contains the full propagation of the constituents and is the object entering momentum-space projections and light-front distributions.

\subsection{Mellin moments of the PDA from the Faddeev wave function}

The Mellin moments of the baryon PDA provide a convenient and physically transparent characterization of its longitudinal structure. They are defined as
\begin{equation}
\label{eq:MellinMomentDef}
\langle x^{l,m,n} \rangle = \int \mathcal{D}x\, x_1^l\, x_2^m\, x_3^n\, \varphi(x_1,x_2,x_3),
\end{equation}
with the integration measure enforcing longitudinal momentum conservation. Our normalization is such that $\langle x^{0,0,0} \rangle = 1$. As a direct consequence of the constraint $x_1+x_2+x_3=1$, the Mellin moments are not all independent. For instance,
\begin{equation}
\label{eq:MellinMomentConservation}
1 = \langle x^{1,0,0} \rangle + \langle x^{0,1,0} \rangle + \langle x^{0,0,1} \rangle \,.
\end{equation}
Thus, only two indices are required to specify an independent set of moments.

Mellin moments can be related to matrix elements of local operators through the operator-product expansion (OPE)~\cite{Lepage:1980fj,Braun:2006hz}. This feature is particularly useful in Euclidean-space approaches, where light-like non-local operators are not directly accessible. In the present framework, Mellin moments are computed from suitable projections of the nucleon Faddeev wave function and subsequently used to reconstruct the proton PDA.

Using Eq.~\eqref{eq:PhiasMatrix1}, one can relate the Mellin moments to projections of the Faddeev wave function. Explicitly, one obtains
\begin{align}
\label{eq:ProjectionMM}
-&\frac{f_N}{2}p\cdot n \slashed{n}N^{\uparrow} \langle x^{m,n} \rangle =
\int \mathcal{D}x\, x_1^m x_2^n 
\int \frac{\textrm{d}^4K}{(2\pi)^4} \frac{\textrm{d}^4k}{(2\pi)^4} \nonumber \\
&\left(\prod_{i=1}^3\delta(p_i-x_i P\cdot n) \right)
\chi^{\alpha_1\alpha_2\alpha_3}_{uud}\,
O^{21}_{\alpha_2\alpha_1}\,
O^3_{\alpha_3},
\end{align}
where $k$ is the relative momentum within the diquark, $K$ its total momentum, and $\chi_{uud}$ the Faddeev wave function projected onto the $\ket{uud}$ basis. The delta functions identify the longitudinal momentum fractions with the light-front projections of the quark momenta.

The operators $O^{21}$ and $O^3$ act as projectors onto the leading-twist Dirac structures defining the PDA. For the amplitude $\varphi$, they are given by
\begin{subequations}
\begin{eqnarray}
\label{eq:DefO12}
O^{21}_\varphi & = &\frac{1-\gamma_5}{2}\,C\,\gamma \cdot n \,\frac{1+\gamma_5}{2}, \\
\label{eq:Def03}
O^3_\varphi & = & \gamma \cdot n \,\frac{1+\gamma_5}{2},
\end{eqnarray}
\end{subequations}
which isolate the appropriate spinor components corresponding to leading-twist contributions \cite{Braun:2006hz}. 

A similar construction applies to the tensor PDA $T$, up to an overall normalization factor and with different projectors,
\begin{subequations}
\begin{eqnarray}
O^{21}_T & = &\frac{1+\gamma_5}{2}\,iC\sigma_{\alpha\beta} n^\alpha \frac{1+\gamma_5}{2} \eta_\perp^{\beta \gamma}, \label{eq:DefO12T}\\
O^3_T & = & \gamma_\delta \gamma \cdot n \frac{1-\gamma_5}{2},
\end{eqnarray}
\end{subequations}
which encode the distinct Dirac structure of this distribution amplitude.

\begin{figure}[!t]
\centering
\includegraphics[width=\linewidth]{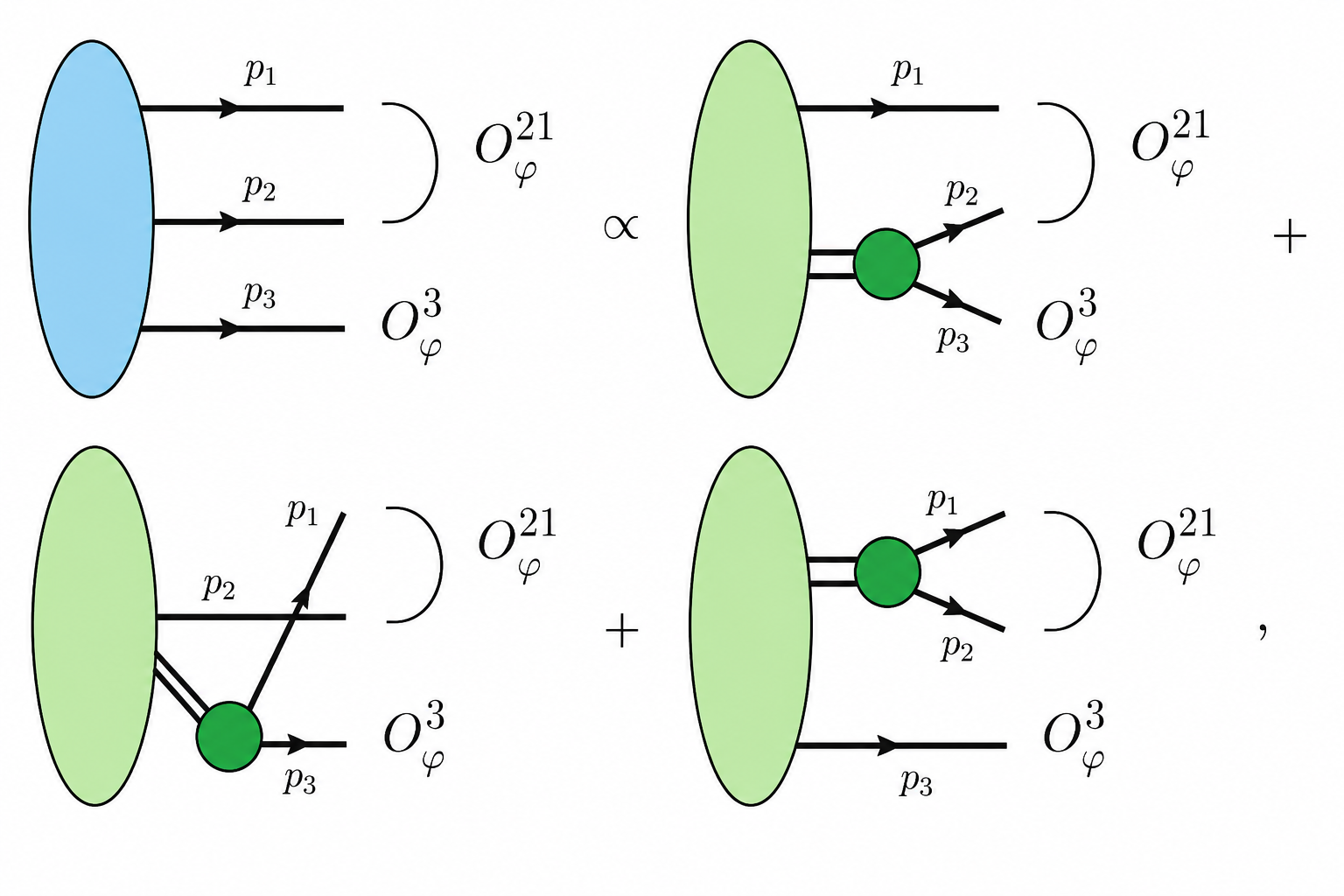}
\caption{\label{fig:Projection} Diagrammatic representation of the Mellin-moment projection in Eq.~\eqref{eq:ProjectionMM}. The blue oval denotes the nucleon Faddeev amplitude, while the green oval indicates a generic scalar- or axial-vector-diquark contribution. Single and double lines represent dressed-quark and diquark propagators, respectively.}
\end{figure}

Equation~\eqref{eq:ProjectionMM} makes explicit that the Mellin moments are obtained from weighted projections of the Faddeev wave function onto specific Dirac and momentum configurations. In the quark--diquark approximation, this projection naturally decomposes into contributions associated with each of the three Faddeev components, $\psi_{1,2,3}$. The corresponding diagrammatic representation is shown in Fig.~\ref{fig:Projection}, where each term reflects a different choice of the bystander quark.

This formulation provides a direct bridge between the covariant description of the baryon in terms of its Faddeev wave function and its partonic structure encoded in the PDA. In particular, it allows one to investigate how different dynamical assumptions on the quark--diquark kernel and the internal structure of the Faddeev amplitude impact the resulting distribution amplitude.

\section{Scalar diquark contribution}
\label{sec:scalardiquarkonly} 

The diagrammatic representation in Fig.~\ref{fig:Projection} indicates that, in principle, three distinct contributions associated with the Faddeev components $\psi_{1,2,3}$ must be considered. However, a substantial simplification arises when restricting the analysis to scalar diquark correlations. 

Considering only scalar diquark correlations inside the proton, two of the three contributions shown in Fig.~\ref{fig:Projection} vanish. The bottom-right diagram yields zero because the projection onto the $\ket{uud}$ state enforces the diquark to be in an isovector configuration ($uu$), whereas scalar diquarks are necessarily isoscalar due to Fermi statistics and colour antisymmetry requirements~\cite{Cahill:1988dx, Cloet:2013jya}.

The bottom-left diagram of Fig.~\ref{fig:Projection} does not contribute at leading twist. The corresponding projection places the two quarks forming the diquark in the same helicity state, whereas a scalar diquark requires a spin-singlet configuration. Such a mismatch can only be compensated by nonzero orbital angular momentum (OAM) and therefore contributes beyond leading twist~ \cite{Lepage:1980fj, Braun:2006hz}.

As a result, only the top-right diagram in Fig.~\ref{fig:Projection} survives. Focusing on this contribution, we introduce the quantity $\mathcal{L}_\nu^{0^+}$, defined through
\begin{equation}
\label{eq:MathLDef}
\gamma \cdot \mathcal{L}^{0^+} =
\left. \chi_{uud}^{\alpha_1\alpha_2\alpha_3}\right|^{0^+} O^{21}_{\alpha_2\alpha_1} O^3_{\alpha_3},
\end{equation}
which isolates the scalar diquark ($0^+$) component of the Faddeev wave function projected onto the leading-twist Dirac structure.

Within the quark--diquark framework, $\mathcal{L}_\nu^{0^+}$ can be written as
\begin{align}
\label{eq:MathL}
\mathcal{L}_\nu^{0^+} &= \frac{1}{4}\textrm{Tr}\Big[ \gamma_\nu  O^{3}_\varphi S_d(p_3) \Gamma^0(k,K)\tilde{S}_u(p_2)\Big] \nonumber \\
&
\times \textrm{Tr}\Big[O^{21}_\varphi S_u(p_1) \Delta^0(K) \mathcal{S}(\ell,P) \Big],
\end{align}
where the various ingredients correspond to the light-dressed-quark propagator, the Bethe--Salpeter amplitude and propagator of a scalar diquark, and the quark--diquark amplitude of the nucleon.

To proceed analytically, we adopt the general decomposition of the dressed-quark propagator,
\begin{equation}
\label{eq:Propagator}
S(p) = -i \gamma \cdot p \,\sigma_A(p^2) + \sigma_B(p^2),
\end{equation}
which is consistent with solutions of the quark Dyson--Schwinger equation \cite{Roberts:1994dr, Cloet:2013jya}. In addition, we retain only the dominant Dirac component of the nucleon amplitude, approximating
\begin{equation}
\label{eq:S1Def}
\mathcal{S}(\ell,P) = s_1(\ell,P)\, I_D,
\end{equation}
which corresponds to neglecting subleading components associated with lower Dirac structures. This approximation is justified at leading twist, where the dominant contribution arises from the $S$-wave component of the nucleon~\cite{Eichmann:2016yit}.

Using cyclic permutations of the trace and standard $\gamma$-matrix identities, one can reorganize Eq.~\eqref{eq:MathL} into
\begin{align}
\label{eq:IntermediateStepMathL}
&
\mathcal{L}_\nu^{0^+} = \frac{1}{4}\textrm{Tr}\Big[ S_d(p_3) \Gamma^0(k,K) \tilde{S}_u(p_2) \frac{1\!-\!\gamma_5}{2}C \gamma\cdot n \nonumber \\
&
\times \frac{1\!+\!\gamma_5}{2} (2n^\nu \!-\! n\cdot \gamma \gamma^\nu) \Big] \sigma_B(p_1) \Delta^0(K) s_1(\ell,P).
\end{align}
Now, exploiting the identity $(\gamma\cdot n)^2 = 0$, valid for light-like vectors, the expression simplifies considerably and factorizes as
\begin{align}
\label{eq:MathLFinal}
\mathcal{L}_\nu^{0^+} &= \frac{1}{4} \textrm{Tr}\Big[ S_d(p_3) \Gamma^0(k,K) \tilde{S}_u(p_2) \frac{1-\gamma_5}{2}C \gamma\cdot n \frac{1+\gamma_5}{2} \Big]\nonumber \\
&
\times \textrm{Tr}\Big[ \gamma^\nu n\cdot \gamma \frac{1+\gamma_5}{2} S(p_1) \mathcal{S}(\ell,P) \Delta^0(K) \Big].
\end{align}

Equation~\eqref{eq:MathLFinal} factorizes into a diquark-PDA projection and a quark--diquark projection coupled through the integration over the diquark momentum $K$. Consequently, the proton PDA is obtained as a convolution of the scalar-diquark PDA with the quark--diquark wave function.


\subsection{Scalar diquark PDA}

We define the Mellin moments of the scalar diquark PDA as
\begin{align}
\label{eq:ScalarDiquarkMMDef}
\mathcal{D}_0^m(K^2) &\propto \int_0^1 \mathrm{d}x \; x^m \varphi_0 (x; K^2), \nonumber\\  
&
=\frac{1}{P \cdot n} \int \frac{\mathrm{d}^4 k}{(2\pi)^4} \; \left(\frac{k \cdot n}{P \cdot n}\right)^m \nonumber\\
&
\times \Tr_D \left[ S_d (k) \Gamma^0(k; K) \Tilde{S}_u (K - k ) \mathcal{O}_\varphi^{21} \right],
\end{align}
which directly follows from the general projection formula for Mellin moments. This expression shows that the diquark PDA can be interpreted as the projection of its Bethe--Salpeter wave function onto light-front momentum fractions, in complete analogy with mesonic distribution amplitudes \cite{Chang:2013pq, Mezrag:2014tva}. In particular, the integration over the internal momentum $k$ scans all possible momentum configurations inside the diquark, while the factor $(k\cdot n)^m$ selects the longitudinal momentum fraction carried by one of its constituents.

To proceed analytically, we adopt an algebraic model for the building blocks entering Eq.~\eqref{eq:ScalarDiquarkMMDef}. The dressed-quark propagator is taken as
\begin{equation}
\label{eq:QuarkPropagator}
S(p) = \frac{-i\gamma \cdot p + M}{p^2 + M^2},
\end{equation}
which captures the essential features of dynamical chiral symmetry breaking while remaining tractable.

For the scalar diquark Bethe--Salpeter amplitude, we consider two scenarios. The first corresponds to a point-like correlation,
\begin{equation}
\label{eq:PointLikeBSA}
\Gamma^0_{\text{PL}}(q,K) C^\dagger = i\gamma_5,
\end{equation}
which provides a useful reference case, with no internal structure. The second, more realistic case, employs a Nakanishi-like representation,
\begin{equation}
\label{eq:ScalarDiquarkBSA}
\Gamma^0(q,K) C^\dagger = i\gamma_5 \int_{-1}^1 \mathrm{d}z \, \rho(z)\, \Lambda_\Gamma^2 \, \sigma_{\Lambda_\Gamma}(q_{+K}^2),
\end{equation}
where $\rho(z)=\frac{3}{4}(1-z^2)$ is a spectral weight, $\sigma_{\Lambda_\Gamma}(q_{+K}^2)=1/(q_{+K}^2+\Lambda_\Gamma^2)$ with $q_{+K}=q-\frac{1-z}{2}K$ and $\Lambda_\Gamma$ a model parameter. This representation provides a smooth interpolation between simple algebraic models and realistic numerical solutions of the Bethe--Salpeter equation~\cite{Mezrag:2014tva, Salme:2017oge}.

Using these ingredients, the Dirac trace appearing in Eq.~\eqref{eq:ScalarDiquarkMMDef} can be computed explicitly. After exploiting the properties of the charge-conjugation matrix and the chiral projectors, one finds that the trace simplifies dramatically and reduces to a scalar structure proportional to $n\cdot K$. This reflects the fact that, at leading twist, only the longitudinal momentum flow survives the projection. Besides, making use of standard $\gamma$-matrix identities, the trace further reduces to
\begin{equation}
\label{eq:TraceReduced}
\mathrm{Tr}_D[\cdots]
=  \int_{-1}^1 
\frac{2M\rho(z)\Lambda_\Gamma^2 (n\cdot K)\,\mathrm{d}z}
{(k^2+M^2)[K_-^2+M^2](k_{+K}^2+\Lambda_\Gamma^2)},
\end{equation}
where $K_-=K-k$. 

The resulting Eq.~\eqref{eq:TraceReduced} involves three propagator denominators, which can be combined using Feynman parameters. Introducing $(u,v,y)$ such that $u+v+y=1$, one obtains a single denominator of the form
\begin{equation}
\int_0^1\frac{\delta(1-u-v-y)\,\Gamma(3)\mathrm{d}u\,\mathrm{d}v\,\mathrm{d}y}
{\left[u(k^2\!+\!M^2)+v(k_{+K}^2\!+\!\Lambda_\Gamma^2)+y(K_-^2\!+\!M^2)\right]^3}.
\end{equation}

After expanding and completing the square, one obtains
\begin{equation}
\int_{0}^1 \mathrm{d}u \mathrm{d}v \mathrm{d}y \; \frac{\delta(1-u-v-y)\Gamma(3)}{\left[(k -\beta K)^2 + \mathcal{M}^2 \right]^3},
\end{equation}
where $\beta= y + \frac{1-z}{2} v$ and $\mathcal{M}^2$ is an effective mass depending on $K^2$, $M$ and $\Lambda_\Gamma$,
\begin{equation}
\mathcal{M}^2 = K^2 \left( y \!+\! \frac{1}{4}(1-z)^2 v \!-\! \beta^2 \right) \!+\! M^2(u+y) \!+\! v\Lambda_{\Gamma}^2.
\end{equation}

Completing the square introduces $k'=k-\beta K$. Since the diquark momentum $K$ inherits a complex temporal component from the on-shell nucleon momentum, this change of variables also displaces the temporal integration contour. The resulting integral must therefore be evaluated using complex-contour methods~\cite{Eichmann:2019dts,Bhagwat:2002tx}. As detailed in Appendix~\ref{app:diquarkContour}, the conditions $M,\Lambda_\Gamma>M_N/3$, satisfied by the scalar-sector parameters employed here, ensure that no poles lie between the displaced contour and the real axis. With the convergence prescription discussed there, the contour can be deformed onto the real axis without additional residue contributions, reducing the calculation to standard Euclidean loop integrals.
The numerator is then expanded in powers of $k'\cdot n$. Odd powers vanish under symmetric integration, whereas positive even powers vanish
upon rotationally invariant tensor reduction because $n^2=0$; for instance, $(k'\cdot n)^2\to k'^2n^2/4$ under the integral. Consequently, only $\beta^m(K\cdot n)^m$ survives from $[(k'+\beta K)\cdot n]^m$. The remaining scalar loop integrals can therefore be evaluated analytically, leaving the integrations over the Feynman and spectral parameters.



The remaining integrations involve the Feynman parameters $(u,v,y)$ subject to the constraint $u+v+y=1$. It is convenient to eliminate one variable using the delta function and to introduce the complementary variable $\bar{u} = 1-u$ and $\Bar{\beta} = y + \frac{1-z}{2}(\Bar{u}-y)$. In this representation, the momentum sharing between the two quarks inside the diquark is encoded in the variables $(\bar{u},y)$.

In order to perform the integration over the spectral parameter $z$, it is advantageous to invert the relation defining $\bar{\beta}$ and express $z$ in terms of $(\bar{u},y,\bar{\beta})$, with the change of variable $z = 2 (\Bar{u} -\Bar{\beta})/(\Bar{u}-y)-1$. This transformation reorganises the integration domain and leads to a representation where the spectral weight $\rho(z)$ becomes a function of $\bar{\beta}$. As a result, the Mellin moments acquire the structure of an integral over $\bar{\beta}$, which can be directly interpreted as the longitudinal momentum fraction inside the diquark.

Finally, by reversing the order of integration, one obtains a compact expression in which the limits of integration explicitly reflect the kinematic constraints $0 \leq y \leq \bar{u} \leq 1$ and $0 \leq \bar{\beta} \leq 1$. In this form, the Mellin moments are written as a convolution over $\bar{\beta}$, weighted by the spectral function and an effective mass scale $\mathcal{M}^2$ that encodes the dynamical content of the diquark.

Collecting all contributions, one arrives at a compact expression for the Mellin moments,
\begin{align}\label{eq.MMs.Scalar.Diquark}
n_0  D_0^m(K^2) &= \frac{4M}{(4\pi)^2}\left( \frac{K\cdot n}{P \cdot n}\right)^{m+1} \nonumber \\
&
\times \int_{0}^1 \mathrm{d}\bar{\beta} \int_{0}^{\bar{\beta} }\mathrm{d}y\int_{\bar{\beta}}^{1}\mathrm{d}\Bar{u}\frac{ \rho(\bar{\beta}) \Lambda_{\Gamma}^2\bar{\beta}^m}{w \left( K^2 + \mathcal{M}^2 \right)},
\end{align}
where the normalisation constant is given by $D_0^0 (K^2 = M_0^2) = n \cdot K / n \cdot P$, with $M_0$ the mass-scale associated with the scalar diquark. 

The Mellin moments defined in Eq.~\eqref{eq.MMs.Scalar.Diquark} correspond to the moments of a function of the longitudinal momentum fraction, so that
\begin{equation}
D_0^m(K^2) \sim \int_0^1 \mathrm{d}x \, x^m \, \varphi_0(x,K^2).
\end{equation}
This relation can be inverted by identifying the dependence on the variable $\bar{\beta}$ in Eq.~\eqref{eq.MMs.Scalar.Diquark} with the longitudinal momentum fraction $x$. Consequently, the PDA is obtained by inspection of the integrand, replacing $\bar{\beta} \to x$ and keeping the remaining integrations as convolutions over the internal degrees of freedom. In this way, the scalar diquark PDA reads

\begin{align}\label{PDA_scalar_diquark0}
 n_0'  \varphi_0(x,K^2)&=  \int_{0}^x \mathrm{d}y \int_{x}^{1}\mathrm{d}\Bar{u} \; \frac{ \rho\left[z(\Bar{u},y,x)\right]\Lambda_{\Gamma}^2}{w' \left( K^2 + \mathcal{M'}^2 \right)}
\end{align}
with the re-definitions
\begin{subequations}
\begin{align}
\mathcal{M'}^2 &= \frac{M^2(1\!-\!\Bar{u}+y) \!+\! \Lambda_{\Gamma}^2 (\Bar{u}\!-\!y)}{w'} (\Bar{u}\!-\!y), \\
w' &= \Bar{u}(y- x^2) +x \left[ y(x-2) + x\right],\\
\rho\left[z(\Bar{u},y,x)\right] &=  \frac{3(\Bar{u}-x)(x-y)}{(\Bar{u}-y)^2},
\end{align}
\end{subequations}
and the normalization constant $n_0'(K^2)$ ensuring that
\begin{equation}
\label{eq:NormalisationPDAScalarDiquark}
\int_0^1 \textrm{d}\hat{x} \varphi_0(\hat{x},1-\hat{x}) = 1 \,.
\end{equation}

The present calculation is restricted to leading twist and retains only the dominant Dirac component of the nucleon Faddeev amplitude. Consequently, orbital-angular-momentum effects and subleading spin-orbit correlations are neglected. In addition, the use of algebraic Ans\"atze for the propagators and diquark amplitudes introduces a degree of model dependence. These approximations are expected to affect the quantitative details of the PDA while preserving its leading qualitative features.

In order to gain further insight into the structure of the scalar diquark PDA, it is useful to consider a simplified scenario in which the model parameters are identified, namely $M=\Lambda_\Gamma$. Under this assumption, the integrations over the Feynman parameters $y$ and $\bar{u}$ in Eq.~\eqref{PDA_scalar_diquark0} can be performed explicitly. After some algebra, one obtains the following closed expression:
\begin{equation}
\varphi_0(x,\lambda) = \lambda \left[1 - \frac{\lambda}{x(1-x)} \ln\left(1 + \frac{x(1-x)}{\lambda}\right)\right],
\end{equation}
where we have introduced the dimensionless parameter
\begin{equation}
\lambda = \frac{M^2}{K^2}.
\end{equation}

At this point, it is convenient to reinterpret the longitudinal momentum fractions in terms of the internal structure of the diquark. Instead of working with $(x_2,x_3)$, defined with respect to the parent baryon, we introduce
\begin{equation}
(\hat{x}_2,\hat{x}_3) = \left(\frac{x_2}{x_2+x_3},\frac{x_3}{x_2+x_3}\right),
\end{equation}
so that $\hat{x}_2+\hat{x}_3=1$. This change of variables allows one to interpret the result as a genuine two-body distribution amplitude, describing how the total diquark momentum is shared between its two constituent quarks.

With this identification, the scalar diquark PDA can be expressed as
\begin{equation}
\label{eq:ScalarDiquarkPDA}
n_0''(K^2) \varphi_0(\hat{x}_2,\hat{x}_3) =\lambda\left[ 1-\frac{\ln\left(1+\hat{x}_2\hat{x}_3/\lambda\right)}{\lambda\hat{x}_2\hat{x}_3}\right],
\end{equation}
where $n_0''(K^2)$ is fixed by the normalisation condition
\begin{equation}
\int_0^1 \mathrm{d}\hat{x} \, \varphi_0(\hat{x},1-\hat{x}) = 1.
\end{equation}

This form makes explicit that the PDA depends both on the internal momentum sharing, encoded in $\hat{x}_2\hat{x}_3$, and on the diquark virtuality $K^2$. The latter controls the degree of correlation between the quarks forming the diquark.

Insight into the structure of the PDA can be obtained by analysing its limiting behaviour. In the regime $K^2 \ll M^2$ (\emph{i.e.} $\lambda \gg 1$), one finds
\begin{equation}
\varphi_0(\hat{x}_2,\hat{x}_3) \approx 6\,\hat{x}_2\hat{x}_3,
\end{equation}
which corresponds to the asymptotic distribution amplitude of a weakly correlated two-body system. In contrast, for $K^2 \gg M^2$ (\emph{i.e.} $\lambda \ll 1$), the PDA tends towards a constant, as expected for a point-like system with maximal correlations. For intermediate values of $K^2$, the PDA remains peaked at $\hat{x}_2=\hat{x}_3=1/2$ but is broader than the asymptotic form, in agreement with studies of light-quark meson PDAs \cite{Chang:2013pq, Qin:2017lcd}. Moreover, its endpoint behaviour is linear, which can be phenomenologically relevant for processes sensitive to the edges of the momentum distribution.

\begin{figure}
\centering
\includegraphics[clip, width=0.45\textwidth]{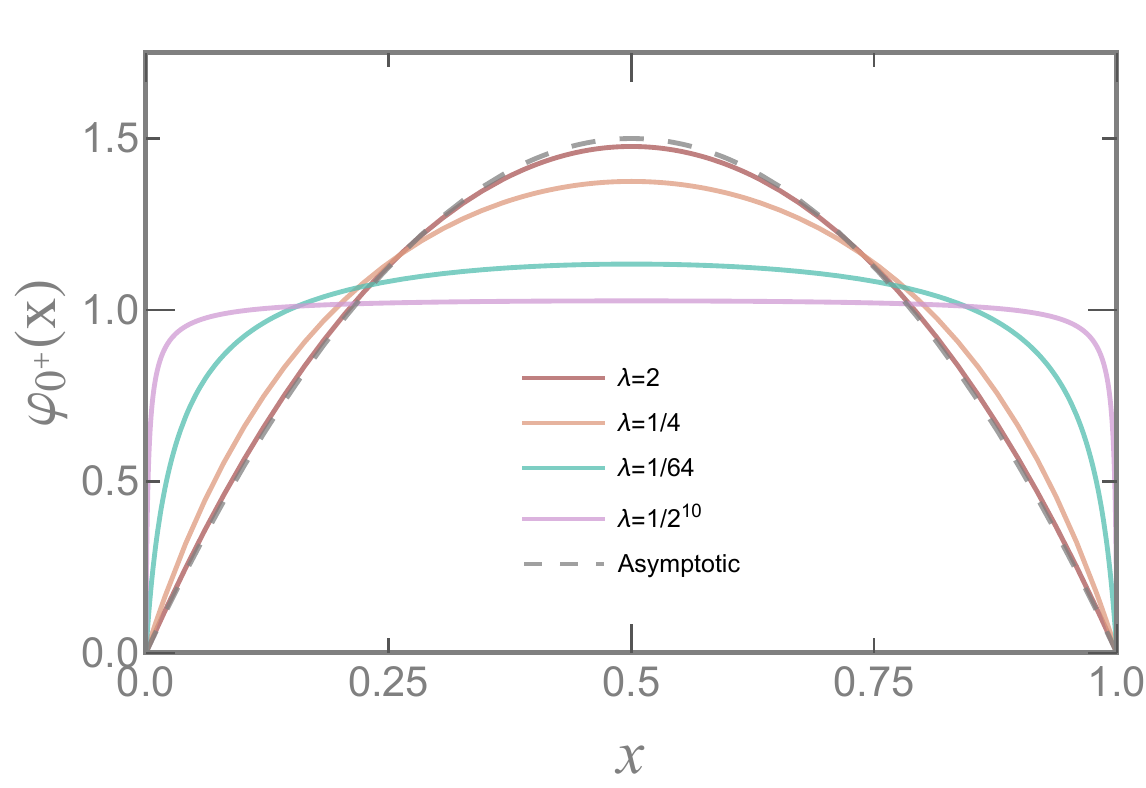}
\caption{\label{fig:ScalarDiquarkPDA} Scalar diquark PDA for representative values of the parameter $\lambda = M^2/K^2$. The curves correspond to: $\lambda=2$ (dark pink), $\lambda=1/4$ (peach tone), $\lambda=1/64$ (turquoise), and $\lambda=1/2^{10}$ (light purple). The asymptotic distribution $\varphi(x)=6x(1-x)$ is shown by the dashed gray line. As $\lambda$ decreases (i.e.\ increasing $K^2$), the PDA evolves from a broad distribution toward a flatter profile, signalling the transition from weakly to strongly correlated quark configurations inside the diquark.}
\end{figure}

The qualitative behaviour of the scalar diquark PDA is illustrated in Fig.~\ref{fig:ScalarDiquarkPDA}, where we display the distribution for representative values of the parameter $\lambda = M^2/K^2$, namely $\lambda=2$, $\lambda=1/4$, $\lambda=1/64$ and $\lambda=1/2^{10}$, together with the asymptotic distribution $\varphi(x)=6x(1-x)$. These cases span a wide range of diquark virtualities, allowing one to visualise the transition from a broad, weakly correlated distribution to a nearly flat profile characteristic of a highly correlated, point-like system.

\subsection{Proton PDA in the scalar diquark approximation}

Restricting ourselves to scalar diquark correlations inside the nucleon, the proton's PDA Mellin moments can be written as
\begin{align}
2 i f_B &\slashed{n} B_{\uparrow} \expval{x_1^l x_2^m}=\int \frac{d^4 K}{(2 \pi)^4}\left(\frac{p_1 \cdot n}{P \cdot n}\right)^l \gamma_\nu \nonumber \\
&
\times \Tr\left[\gamma_\nu O_\varphi^3 S_u\left(p_1\right) \mathcal{S}(l;P) \Delta^{0^+}(K)\right]D_0^m(K^2),
\end{align}
where the scalar diquark propagator is parametrised as
\begin{equation}
    \Delta^{0^+}(K)=\sigma_{M_0}\left(K^2\right)=\frac{1}{K^2+M_0^2}.
\end{equation}

In this expression, the Mellin moments of the scalar diquark PDA, $D_0^m(K^2)$, appear naturally as building blocks of the proton distribution amplitude. Physically, this reflects the hierarchical structure of the problem: the proton PDA is obtained by combining the internal momentum distribution of the scalar diquark with the relative momentum distribution between the bystander quark and the diquark correlation.

The quark--diquark interaction is encoded in the scalar component of the nucleon Faddeev amplitude,
\begin{align}
n \mathcal{S}(l=-K + 2 P / 3; P)&=i \int_{-1}^1 \mathrm{d}z \; \rho(z) \hat{\sigma}_{\Lambda_{0p}} \left(K_{+P}^2\right)
\end{align}
where the spectral weight $\rho(z)$ controls the momentum distribution inside the quark--diquark system and $\hat{\sigma}_{\Lambda_{0p}}(K_{+P}^2) = \Lambda_{0p}^2 \, \sigma_{\Lambda_{0p}}(K_{+P}^2)$. The algebraic form employed here allows the resulting momentum integrals to be evaluated analytically while preserving the dominant non-perturbative features of the bound state.

Substituting the explicit expressions for the propagators and amplitudes into the Mellin moments equation yields
\begin{align}
&2 i f_B \slashed{n} B_{\uparrow} \expval{x_1^l x_2^m}= \frac{i\Lambda_{0p}^{2}}{2} \int \frac{\mathrm{d}^4 K}{(2 \pi)^4}\left(\frac{p_1 \cdot n}{P \cdot n}\right)^l \int_{-1}^1 \mathrm{d}z\nonumber \\
& \frac{\frac{3}{4}(1-z^2)\gamma_\nu \Tr\left[\gamma_\nu\slashed{n}\left(i+\gamma_5\right)\left(-\slashed{p}_1+M\right)\right]D_0^m(K^2)}{\left(p_1^2+M^2\right)\left(K_{+P}^2+\Lambda_{0p}^{2}\right)\left(K^2+M_0^2\right)} .
\end{align}

The Dirac trace can be evaluated straightforwardly,
\begin{align}
\gamma_\nu \Tr\left[\gamma_\nu\slashed{n}\left(i+\gamma_5\right)\left(-\slashed{p}_1+M\right)\right]=4 M \slashed{n},
\end{align}
showing explicitly that only the scalar component of the dressed-quark propagator contributes at leading twist. This is consistent with the fact that the leading-twist PDA projects the dominant light-front configuration of the nucleon.

Using the expression previously obtained for the scalar diquark Mellin moments, Eq. \eqref{eq.MMs.Scalar.Diquark}, and simplifying, one arrives at
\begin{align}
f_B  B_{\uparrow} &\expval{x_1^l x_2^m} =  \frac{3M^2\Lambda_{0p}^{2} \Lambda_\Gamma^2}{(4\pi)^2} \int \frac{\mathrm{d}^4 K}{(2 \pi)^4}\left(\frac{p_1 \!\cdot \! n}{P \! \cdot \! n}\right)^l \left(\frac{K \!\cdot\! n}{P \!\cdot \!n}\right)^{m+1} \nonumber \\
&
\times \int_{-1}^1 \mathrm{d}z \left(1-z^2\right) \int_0^1 \mathrm{d}\beta \beta^m \int_0^\beta \mathrm{d}y \int_\beta^1 \mathrm{d}\Bar{u} \nonumber \\
&
\frac{\rho \left[ z(\bar{u},y,\beta)\right]}{w(K^2\!+\!\mathcal{M}^2)\left(p_1^2\!+\!M^2\right)\left(K_{+P}^2\!+\!\Lambda_{0p}^{2}\right)\left(K^2\!+\!M_0^2\right)},
\end{align}
with
\begin{subequations}
\begin{align}
\rho(z(\beta, \bar{u}, y))& =3 \frac{(u-\beta)(\beta-y)}{(\bar{u}-y)^2}, \\
w &= \Bar{u}(y - \beta^2) + \beta (y(\beta-2)+\beta).
\end{align}
\end{subequations}

The calculation involves now four propagator denominators. In order to combine them into a single quadratic form suitable for momentum integration, four Feynman parameters are introduced:
\begin{align}
&\frac{1}{(K^2+\mathcal{M}^2)\left(p_1^2+M^2\right)\left(K_{+P}^2+\Lambda_{0p}^{2}\right)\left(K^2+M_0^2\right)} = \nonumber \\ 
&\Gamma(4) \int \frac{\mathrm{d}a_1 \mathrm{d}a_2 \mathrm{d}a_3 \mathrm{d}a_4 \; \delta\left(1-a_1-a_2-a_3-a_4\right)}{D_1^4},
\end{align}
where the denominator is
\begin{align}
D_1 &=\left[\left((K-P)^2+M^2\right) a_1 + \left(K_{+P}^2-\Lambda_{0p}^{2}\right) a_2 \right. \nonumber \\
&
\left.+\left(K^2+M_0^2\right) a_3+\left(K^2+\mathcal{M}^2\right) a_4\right].
\end{align}

After combining the denominators and completing the
square, we introduce $K'=K-\alpha P$, where $\alpha=a_1+\frac{1-z}{2}a_2$. Since the on-shell nucleon momentum has a complex Euclidean temporal component, this change of variables also displaces the temporal integration contour. Its deformation onto the real axis must therefore be justified using Cauchy's theorem, accounting for any crossed poles and checking that the contributions at infinity vanish.
Using $p_1=P-K$, the longitudinal momentum factors become
\begin{align}
&\left(\frac{p_1\cdot n}{P\cdot n}\right)^l
 \left(\frac{K\cdot n}{P\cdot n}\right)^{m+1}
\nonumber\\
&\quad =
\left(1-\alpha-\frac{K'\cdot n}{P\cdot n}\right)^l
\left(\alpha+\frac{K'\cdot n}{P\cdot n}\right)^{m+1}.
\end{align}
For the resulting rotationally invariant Euclidean integral, odd powers of $K'\cdot n$ vanish by symmetric integration, while positive even powers vanish upon tensor reduction because $n^2=0$. Thus, within this integral, only the term $(1-\alpha)^l\alpha^{m+1}$ survives.


After performing the momentum integration and implementing appropriate rescalings of the Feynman parameters,
\begin{equation}
\beta'=\alpha\beta,
\qquad
\bar{u}'=\alpha\bar{u},
\qquad
y'=\alpha y,
\end{equation}
the Mellin moments take the form
\begin{align}
f_B  B_{\uparrow} & \expval{x_1^l x_2^m} =  \frac{8M^2\Lambda_{0p}^{2} \Lambda_\Gamma^2}{(4\pi)^4}  \int_{0}^{1} \mathrm{d}\alpha\left(1-\alpha\right)^l \int_0^\alpha \mathrm{d}\beta' \beta'^m  \nonumber \\
&
\times \int_0^\alpha \mathrm{d}a_1 \int_{\alpha}^{1} \frac{\mathrm{d}a_2'}{a_2' - a_1} \int_0^{1-a2'} \mathrm{d}a_3  \nonumber \\
&
\times \int_0^{\beta'}\mathrm{d}y' \int_{\beta'}^\alpha \mathrm{d}\Bar{u}' \frac{\alpha \rho(z)\rho \left[ z(\bar{u}',y',\beta')\right]}{d_1 d_2^2}.
\end{align}

The final structure of this expression makes the identification of the proton PDA immediate. Indeed, comparing with the Mellin moment definition,
\begin{equation}
\label{eq:FinalScalarMMFormula}
\expval{x_1^l x_2^m} = \int_0^1 \mathrm{d}\alpha (1-\alpha)^l\int_0^\alpha \mathrm{d}\beta\; \beta^m \psi_{0^+}(1-\alpha,\beta),
\end{equation}
one identifies the scalar contribution to the proton PDA as
\begin{align}
\mathcal{N}_{0^+} & \psi_{0^+}(1-\alpha,\beta)= \int_0^\alpha \mathrm{d}a_1 \int_{\alpha}^{1} \frac{\mathrm{d}a_2'}{a_2' - a_1} \int_0^{1-a2'} \mathrm{d}a_3 \nonumber \\
&
\int_0^{\beta'}\mathrm{d}y' \int_{\beta'}^\alpha \mathrm{d}\Bar{u}' \, \frac{\alpha \rho(z)\rho \left[ z(\bar{u}',y',\beta')\right]}{d_1 d_2^2} \,,
\end{align}
where
\begin{subequations}
    \begin{align}
d_1=& \bar{u}'(\alpha y'-\beta'^2)+ \beta' \left( y'(\beta' -2\alpha)+\beta' \alpha \right),\\
d_2 =&P^2\left( a_1 + \frac{1}{4}(1-z)^2 a_2 - \alpha^2 \right) +M^2 a_1+\Lambda_{0p}^{2} a_2 \nonumber \\
&+M_0^2 a_3+\mathcal{M}^2 (1-a_1 - a_2 - a_3).
\end{align}
\end{subequations}

The physical interpretation of the resulting variables becomes transparent at this stage. The quantity $x_1=1-\alpha$ represents the longitudinal momentum fraction carried by the bystander quark, whereas the momentum fractions associated with the quarks inside the scalar diquark correlation are encoded through $\beta'$ and $x_3=1-x_1-x_2$. Consequently, the full proton PDA describes the sharing of longitudinal momentum between the active quark and the correlated quark pair generated dynamically by the Poincaré-covariant Faddeev kernel.

Since Eq.~\eqref{eq:FinalScalarMMFormula} is valid for arbitrary non-negative integers $(l,m)$ and
\begin{align}
\label{eq:MeasureIntegration}
\int_0^1 \mathcal{D}x\, x_1^l x_2^m  & f(x_1,x_2,x_3) = \int_0^1\textrm{d}x_1 \, x_1^l \nonumber \\
&
\int_0^{1-x_1}\textrm{d}x_2\, x_2^m f(x_1,x_2,1-x_1-x_2),
\end{align}
the scalar contribution to the proton PDA can be identified directly from the integrand structure by performing the mapping
\begin{equation}
x_1 = 1-\alpha, \qquad x_2=\beta', \qquad x_3=\alpha-\beta'.
\end{equation}

The remaining integrations over the Feynman parameters are computed numerically using the parameter set displayed in Table~\ref{tab:ParameterScalar}. These values are chosen consistently with previous studies of the nucleon Faddeev amplitude within the quark--diquark approximation~\cite{Cloet:2008re, Segovia:2014aza, Chen:2017pse}. In particular, the scales $M$ and $\Lambda_\Gamma$ characterise the dressed-quark propagator and the width of the diquark Bethe--Salpeter amplitude, respectively, while $M_0$ and $\Lambda_{0p}$ control the propagation and relative momentum distribution of the scalar diquark correlation inside the nucleon.

\begin{figure}[!t]
\centering
\includegraphics[width=0.4\textwidth]{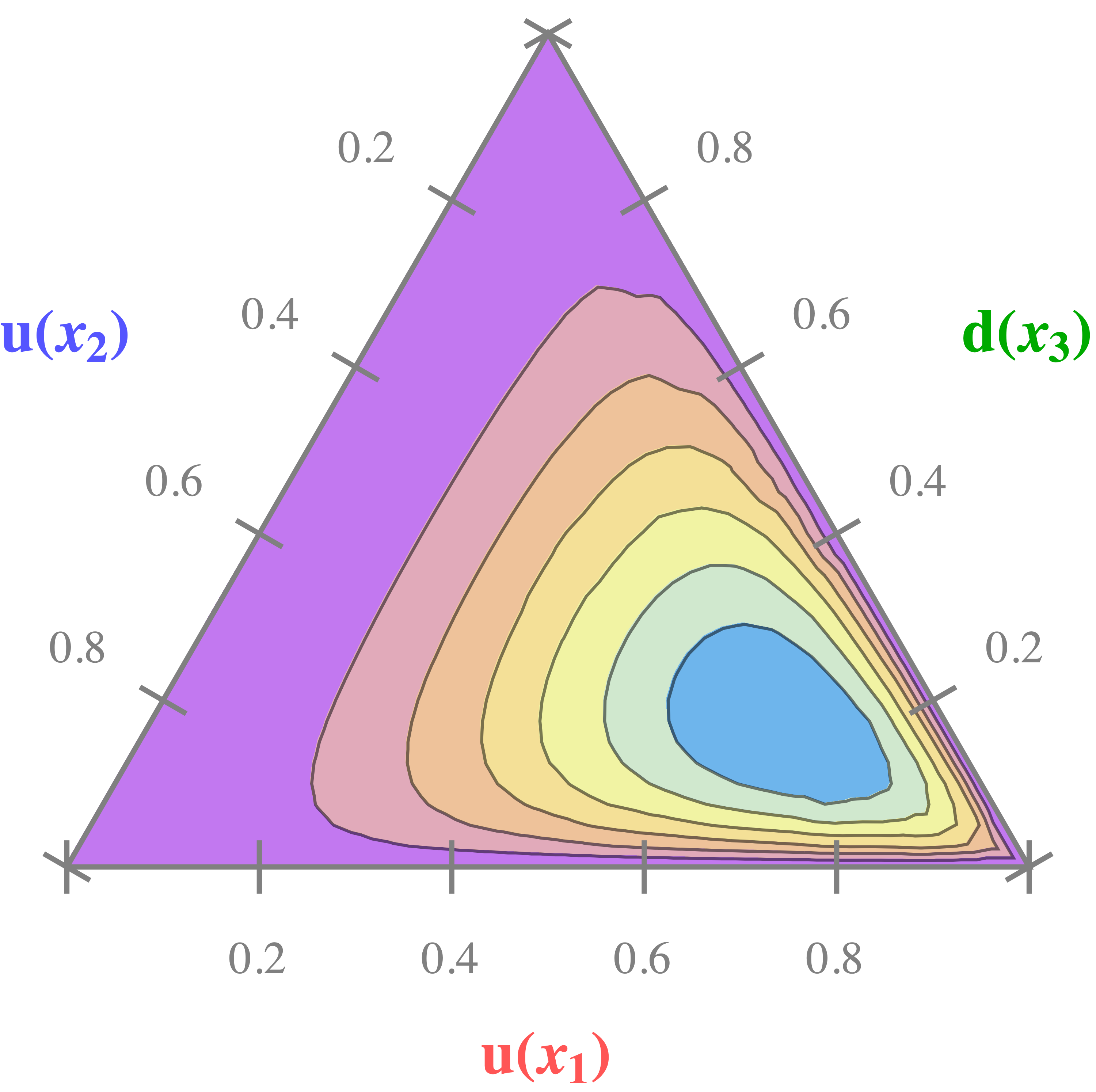}
\includegraphics[width=0.4\textwidth]{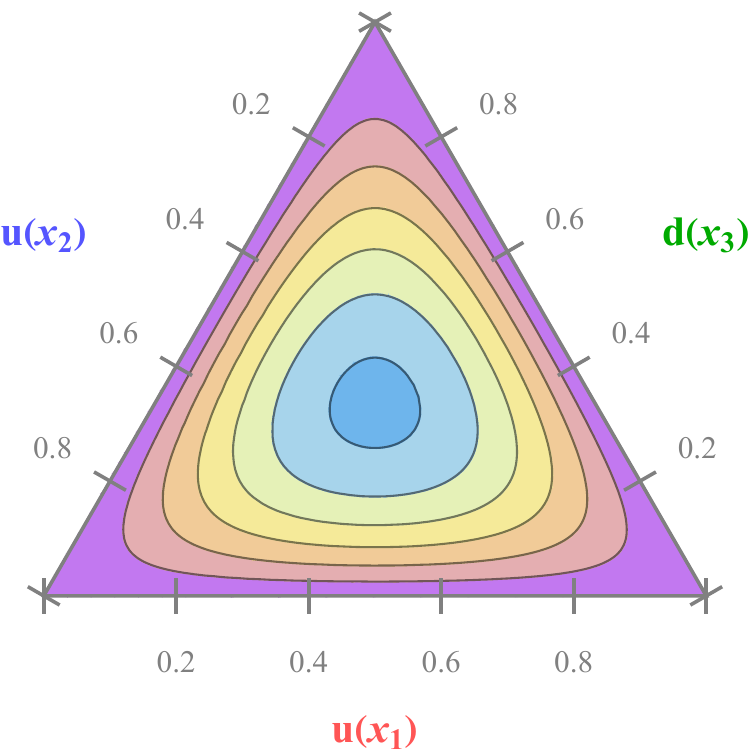}
\caption{\label{fig:NucleonPDAScalarOnly} Proton PDA obtained in the scalar-diquark approximation. Upper panel: Nakanishi-like case; Lower panel: asymptotic distribution. The scalar-diquark result is clearly broader and more asymmetric than the asymptotic PDA, with support shifted toward large values of the spectator-quark momentum fraction. The distribution is displayed in barycentric coordinates $(x_1,x_2,x_3)$, with $x_1+x_2+x_3=1$.}
\end{figure}

\begin{table}[!t]
\centering
\begin{tabular}{cccccc}
\hline\hline
\hspace*{0.1cm} $M$ \hspace*{0.1cm} & \hspace*{0.1cm} $\Lambda_\Gamma$ \hspace*{0.1cm} & \hspace*{0.1cm} $M_0$ \hspace*{0.1cm} & \hspace*{0.1cm} $\Lambda_{0p}$ \hspace*{0.1cm} & \hspace*{0.1cm} $M_1$ \hspace*{0.1cm} & \hspace*{0.1cm} $\Lambda_{1p}$ \hspace*{0.1cm} \\
\hline
$2/5$ & $2/5$ & $2/3$ & $1$ & $3/4$ & $3/5$ \\
\hline\hline
\end{tabular}
\caption{\label{tab:ParameterScalar} Model parameters used to compute the quark--scalar-diquark contribution to the nucleon's PDA. Numbers are given in units of the nucleon mass, $m_p=1$.}
\end{table}

The resulting PDA shown in the upper panel of  Fig.~\ref{fig:NucleonPDAScalarOnly} exhibits two characteristic non-perturbative features. First, it is noticeably broader than the asymptotic distribution (lower panel of Fig.~\ref{fig:NucleonPDAScalarOnly}), reflecting the presence of strong dynamical correlations within the nucleon dressed-quark core~\cite{Lepage:1980fj, Chernyak:1984bm, Cloet:2008re}. Second, the distribution is skewed toward the large-$x_1$ region, indicating that the bystander quark tends to carry a larger fraction of the longitudinal momentum than the quarks participating in the scalar diquark correlation. This behaviour emerges naturally within the quark--diquark picture since the two correlated quarks share momentum inside a composite subsystem, while the remaining quark effectively recoils against the diquark correlation~\cite{Cahill:1988dx, Eichmann:2016yit, Segovia:2015ufa}. Consequently, the PDA departs from the symmetric asymptotic and retains direct information about the internal organisation of momentum generated by the Poincaré-covariant Faddeev dynamics~\cite{Segovia:2013rca, Chen:2017pse}. Such distortions of the distribution amplitude are therefore clear signatures of non-trivial quark--quark correlations in baryons and encode essential information about the non-perturbative structure of the nucleon.


\section{Axial-Vector diquark contribution}
\label{sec:axialvector}

The scalar-diquark contribution derived in the previous section captures an important component of the nucleon Faddeev wave function. Nevertheless, realistic Poincaré-covariant descriptions of the nucleon require the inclusion of axial-vector $(1^+)$ diquark correlations, which play a crucial role in the nucleon's spin-flavour structure and in the description of numerous baryon observables~\cite{Cahill:1987qr, Segovia:2015hra, Eichmann:2016yit, Chen:2017pse, Cloet:2008re}. These degrees of freedom are therefore necessary for constructing the complete leading-twist proton PDA.

The implementation of the axial-vector diquark sector is qualitatively richer than the scalar case because a spin--1 correlation admits different leading-twist projections. This feature is well known from the study of vector mesons where, for instance, the $\rho$ meson possesses two independent leading-twist PDAs associated with different chiral structures~\cite{Chernyak:1983ej, Ball:1996tb}.

Proceeding analogously to the scalar-diquark case, we introduce the projected quantity $\mathcal{L}^{1^+}_{i,\nu}$ through
\begin{equation}
\label{eq:LAVDef}
\sum_{i=1}^3 o^i \gamma \cdot \mathcal{L}_{i}^{1^+} = \left. \chi_{uud}^{\alpha_1\alpha_2\alpha_3}\right|^{1^+} O^{21}_{\alpha_2\alpha_1}O^3_{\alpha_3}, 
\end{equation}
where $i$ labels the bystander quark in the different Faddeev components shown in Fig.~\ref{fig:Projection}, and $o^i$ are the corresponding isospin coefficients. In contrast with the scalar sector, all three contributions survive in the axial-vector channel because the axial-vector diquark belongs to the isotriplet representation and is therefore compatible with both $uu$ and $ud$ flavour combinations.

The quark--axial-vector-diquark amplitude is modelled through
\begin{align}
\label{eq:AmuDef}
\eta A_\mu (\ell = -K+2/3P ; P ) &= \frac{1}{6}r_A \gamma_5\left(\gamma_\mu-i r_PP_\mu \right) \nonumber \\
&
\times \int_{-1}^1\textrm{d}z \Lambda^2_{1p}\sigma_{\Lambda_{1p}}(K_-),
\end{align}
where $K_-=K-\frac{1-z}{2}P$, $\eta$ is a normalization constant, $r_A$ controls the relative weight of scalar and axial-vector diquark contributions, and $r_P$ determines the relative importance of the two Dirac structures entering the axial-vector amplitude. The latter parameter is chosen such that the $P_\mu$ component contributes with relative weight $13/87$, following the structure obtained in numerical solutions of the contact-interaction Faddeev equation~\cite{Chen:2012qr}. Physically, the presence of both $\gamma_\mu$ and $P_\mu$ terms reflects the richer spin structure of the spin--1 diquark correlation and allows the nucleon wave function to incorporate non-trivial spin-orbit dynamics.

Using Eq.~\eqref{eq:AmuDef}, together with the projector structure introduced in Eq.~\eqref{eq:LAVDef}, one obtains
\begin{subequations}
\begin{align}
\mathcal{L}_{1,\nu}^{1^+} &=  \frac{1}{4} \textrm{Tr}\left[S_d(p_3)\Gamma_\mu^1(k,K)S^T_u(p_2)\mathcal{O}_{\varphi}^{21}\right]  \label{eq:L1Result}\nonumber \\
&
\times\textrm{Tr}\left[\gamma_\nu \mathcal{O}_{\varphi}^{3}S_u(p_1)A_\rho \right]\Delta^1_{\mu\rho}(K),\\
\mathcal{L}_{2,\nu}^{1^+} &= -\frac{1}{4}\textrm{Tr}\left[\mathcal{O}_{T}^{21}S_d(p_3)\Gamma_\mu^1(k,K) S_u^T(p_1)C \right]\nonumber \\
&
\times\textrm{Tr}\left[\gamma_\nu\mathcal{O}_{T}^{3} S_u(p_2)A_\rho\right] \Delta^1_{\mu\rho}(K), \label{eq:L2Result}\\
\mathcal{L}_{3,\nu}^{1^+} &= \frac{1}{4}\textrm{Tr}\left[S_u(p_2)\Gamma_\mu^1(k,K)S^T_u(p_1)\mathcal{O}_{\varphi}^{21}\right] \nonumber \\
&
\times\textrm{Tr}\left[\gamma_\nu 
\mathcal{O}_{\varphi}^{3}S_d(p_3)A_\rho \right] \Delta^1_{\mu\rho}(K) \label{eq:L3Result},
\end{align}
\end{subequations}
with the axial-vector diquark propagator given by
\begin{equation}
\label{eq:AVPropagator}
\Delta_{\mu\rho}^1(K) = \left(\delta_{\mu\rho}- \frac{K_\mu K_\rho}{K^2}\right) \sigma_{M_1}(K^2).
\end{equation}
    
Several important observations can already be made at this stage. First, Eqs.~\eqref{eq:L1Result} and \eqref{eq:L3Result} possess identical Dirac structures up to the interchange of the quark labels $u_1 \leftrightarrow d_3$. Their diquark projections involve operators of the form $\gamma\cdot n$, precisely analogous to the chiral-even projection defining the longitudinal $\rho$-meson PDA~\cite{Chernyak:1983ej, Ball:1996tb}. Consequently, these two terms naturally generate convolutions involving the chiral-even axial-vector-diquark PDA.

On the other hand, Eq.~\eqref{eq:L2Result} contains a tensor projection proportional to $\sigma^{\perp\lambda}n_\lambda$, directly mirroring the structure entering the transverse $\rho$-meson PDA~\cite{Chernyak:1983ej, Ball:1996tb}. This implies that $\mathcal{L}^{1+}_{2,\nu}$ probes the chiral-odd component of the axial-vector diquark wave function. Therefore, unlike the scalar sector, the proton's leading-twist PDA receives simultaneous contributions from both chiral-even and chiral-odd axial-vector-diquark distributions.

These observations are physically significant because they demonstrates that the nucleon leading-twist PDA is sensitive not only to momentum correlations but also to the spin structure of the underlying diquark degrees of freedom. The axial-vector sector introduces additional spin-alignment mechanisms and non-trivial helicity structures that are absent in the scalar approximation. As a consequence, the inclusion of axial-vector correlations is expected to modify both the shape and symmetry properties of the final proton PDA, yielding a more realistic representation of the baryon's non-perturbative light-front structure.

\subsection{Axial-vector diquark PDAs}

The evaluation of the axial-vector diquark contribution follows the same general strategy employed in the scalar case. Nevertheless, the richer Dirac structure of the axial-vector correlation introduces two independent leading-twist projections, chiral-even and chiral-odd PDAs, each encoding distinct spin-momentum correlations inside the diquark subsystem.

At the algebraic level, the calculation proceeds through the Mellin moments of the corresponding light-front distributions. The chiral-even projection is obtained from
\begin{equation}
D_{1^+}^m(K^2)\propto \int_0^1 \mathrm{d}x\, x^m \varphi_{1^+}^{\mathrm{even}}(x;K^2),
\end{equation}
with the moments computed from the axial-vector Bethe--Salpeter wave function:
\begin{align}
D_{1^+}^m(K^2)&= \frac{1}{n\!\cdot\! P} \int \frac{\mathrm{d}^4 k}{(2\pi)^4} \left(\frac{n\!\cdot\! k}{n\!\cdot\! P}\right)^m
\nonumber\\
&\times \Tr_D\left[S_d(k)\Gamma^{1j}_\mu(k;K)\widetilde S_u(K-k)
\mathcal O_\varphi^{21}\right].
\end{align}

Using the algebraic \emph{ansatz} for the axial-vector Bethe--Salpeter amplitude,
\begin{align}
n_1\Gamma^{1j}_\mu(k;K)C^\dagger &= io^j \left( \gamma_\mu^T + r_1 \mathfrak{f}(k;K) [\slashed{k},\gamma_\mu^T] \right) \nonumber \\
&
\times \int_{-1}^1\mathrm{d}z\, \rho(z)\, \hat{\sigma}_{\Lambda_\Gamma}(k_{+K}^2),
\end{align}
where $n_1$ is a normalization constant, $\mathfrak{f}(k;K)=k\cdot K/(k^2K^2(k-K)^2)^{1/2}$, $k_{+K} = k + (z-1)K/2$ and $\gamma_{\mu}^T= T_{\mu \nu} (K) \gamma_\nu = \left( \delta_{\mu \nu} - K_\mu K_\nu/K^2 \right)\gamma_\nu$. The isospin coefficients are denoted as $(I=1,I_z=1)=o^+=\sqrt{2}$ and $(I=1,I_z=0)=o^0=-1$. Besides, we define $\hat{\sigma}_{\Lambda_\Gamma}(s) = \Lambda_\Gamma^2/(s+\Lambda_\Gamma^2)$, $k_{+K} = k + (z-1)K/2$, where $(\Lambda_{\Gamma})^{-1}$ measures the diquark size, and $\rho(z)$ as a spectral function. Together with the dressed-quark propagators,
\begin{equation}
S_q(k)=\frac{-i\gamma\!\cdot\!k+M_q}{k^2+M_q^2},
\end{equation}
the Dirac trace can be evaluated analytically.

A noteworthy simplification occurs in the chiral-even channel: the term proportional to $[\slashed{k},\gamma_\mu^T]$ vanishes after the Dirac trace is evaluated. Consequently, only the transverse $\gamma_\mu^T$ component contributes to the longitudinal leading-twist PDA. After introducing Feynman parameters, shifting the loop momentum, and carrying out the Euclidean momentum integration, the Mellin moments reduce to
\begin{align}
\label{MMs_algebraic_axialvectordiquark}
n_{1^+}^{e} D_{1^+}^{m\mu}(K^2) &= n_\nu T_{\mu\nu}(K)\frac{(n\!\cdot\!K)^m}{(n\!\cdot\!P)^{m+1}} \nonumber \\
&
\times \int_0^1\mathrm{d}xx^m \mathcal F_e(x,\lambda),
\end{align}
where $\lambda=M^2/K^2$ and $\mathcal F_e$ contains the remaining integrations over Feynman parameters
\begin{align}
\mathcal F_e(x,\lambda)&= \int_0^{x} \mathrm{d}y \int_{x}^1 \mathrm{d}\Bar{u} \;  \frac{(\Bar{u}-x)(y-x)}{(\Bar{u}-y)^2}\nonumber \\
&\times \frac{ \lambda - x(1-x) }{\Bar{u}(y+\lambda -x^2) +y(x^2 - 2x -\lambda) + x^2}.
\end{align}

The resulting integrations can be performed analytically, yielding the compact expression
\begin{align}
    n_{1^+}^{e}&\varphi_{1^+}^{\mathrm{even}}(x,\lambda)=\nonumber\\
    &\left[x(1\!-\!x)\!+\!\lambda\right]\left[1\!-\!\dfrac{\lambda}{x(1\!-\!x)}\ln\left(1\!+\!\frac{x(1\!-\!x)}{\lambda}\right)\right],
\end{align}
this expression exhibits the expected asymptotic behaviour. In the limit $\lambda\ll1$ (\emph{i.e.} $K^2\gg M^2$), one recovers
\begin{equation}
\varphi_{1^+}^{\mathrm{even}}(x)=6x(1-x),
\end{equation}
which corresponds to the asymptotic PDA.

The chiral-odd projection is algebraically more involved because, contrary to the previous case, it receives contributions exclusively from the commutator structure $[\slashed{k},\gamma_\mu^T]$. The Mellin moments are defined through
\begin{align}
D_{1^+O}^{m\mu}(K^2)&= \frac{1}{n\!\cdot\!P}\int\frac{\mathrm{d}^4k}{(2\pi)^4}
\left(\frac{n\!\cdot\!k}{n\!\cdot\!P}\right)^m \nonumber \\
&
\times \Tr_D \left[S_d(p_3)\Gamma_\mu^{1j}(k;K)\tilde S_u(p_1)
\mathcal O_{T\alpha}^{21}\right].
\end{align}

Because of the additional momentum dependence generated by the tensor projection, the denominator contains five propagator-like factors. Accordingly, five Feynman parameters must be introduced. After performing the parametrization, shifting the loop momentum, and evaluating the Dirac algebra, the Mellin moments can again be written in the standard form
\begin{align}\label{eq6_avodd}
n_{1^+}^{o}D_{1^+ O}^{m\mu}(K^2) &= (\delta_{\mu\beta} \;n \cdot K - n_\mu K_\beta) \delta_{\beta \mu}^\perp\frac{(n \cdot K)^m}{(n \cdot P)^{m+1}} \nonumber \\
&
\times\int_{0}^1 \mathrm{d}x \;x^{m} \mathcal F_o(x,\lambda),
\end{align}
with
\begin{align}
\mathcal{F}_o(x,\lambda) &= \int_x^1 \mathrm{d}b \int_0^{1-b} \mathrm{d}u_2 \int_0^{1-u_2}\mathrm{d}a \int_0^{a} \mathrm{d}v_2 \nonumber \\
&
\times\frac{x\rho\left[z(a,b,x) \right]}{(b-a) (u_2 v_2)^{1/2}} \nonumber \\
&
\times \left[ \frac{x-1}{\mathcal{M}_{o}^{1^+}(K,x)} - \frac{x^2(x-1)+x M^2}{2\left(\mathcal{M}^{1^+}_{o}(K,x)\right)^2} \right],
\end{align}
where 
\begin{align}
\mathcal{M}_{o}^{1^+}(K,x) &= K^2 (a+ \frac{1}{4} y (1-z)^2 - x^2) \nonumber \\
&
+ (u_1 + v_1) M^2 + (b-a) \Lambda_\Gamma^2.
\end{align}

\begin{figure}[t]
\centering
\includegraphics[clip, width=0.45\textwidth]{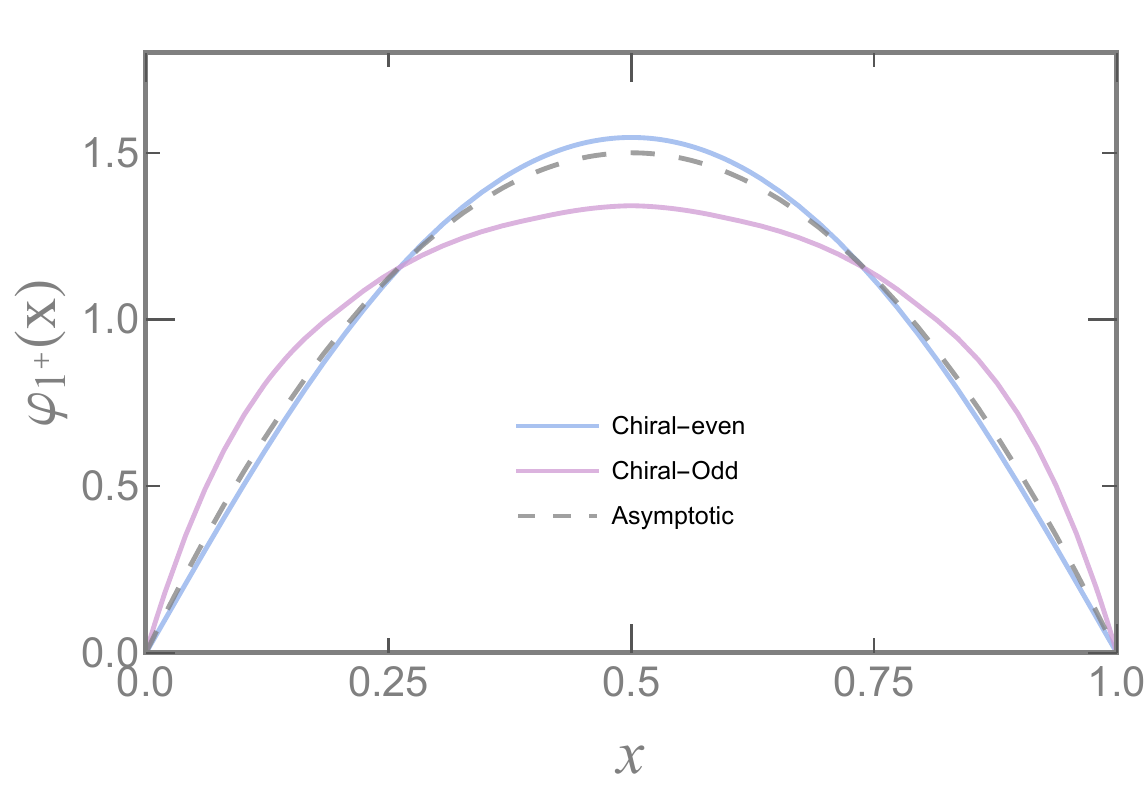}
\caption{\label{fig:AVDiquarkPDA} Parton distribution amplitudes (PDAs) of the axial-vector diquark. \emph{Purple-solid} curve denotes the chiral odd case, the \emph{blue-solid} curve is the chiral even case and the \emph{black-dashed} curve represents the asymptotic PDA, $\varphi(x)=6x(1-x)$.}
\end{figure}

Defining $\lambda=\frac{M^2}{K^2}$ and setting $\Lambda_\Gamma^2=M^2$, the final result can be expressed as
\begin{align}
n_{1^+}^o  \varphi^{\mathrm{odd}}_{1^+}(x,&\lambda,\nu)=(x-1) f(x,\lambda, 1)\nonumber \\
&-\frac{1}{2} x\left[x(x-1) K^2+M^2\right] f(x, \lambda,2),
\end{align}
here, $j=1,2$ is an auxiliary index specifying the denominator power in the remaining Feynman-parameter integral; it is unrelated to the exponent $\nu$ introduced in the Bethe--Salpeter ansatz.
\begin{align}
f(\beta, \lambda, j) &=\int_\beta^1 \mathrm{d}b \int_0^{1-b} \mathrm{d} u_1 \int_0^{1-u_1} \mathrm{d}a \int_0^a \mathrm{d}v_2 \nonumber \\
&
\times\frac{\rho[z(a,b,\beta)]M^2 \beta}{(b-a)\left(u_1 v_1\right)^{1/2}} \frac{1}{\left(\mathcal{M}^{1^+}_{o}(\beta,\lambda)\right)^j}.
\end{align}

To simplify the previous expression we will define the following functions
\begin{align}\label{eq.Iq4}
I_{q^4}  \equiv &(\beta-1) f(\beta, \lambda, 1) \nonumber\\
 =&\int \mathrm{d}a \mathrm{d}b \mathrm{d}v_1 \mathrm{d}u_1 \frac{\rho[z(a, b, \beta)] }{(b-a)\left(u_2 v_2\right)^{1 / 2}} \nonumber\\
&\times \frac{\lambda\beta (\beta-1)}{\left[\lambda\left(1\!-\!u_2\!-\!v_2\right)\!+\!a\!+\!\frac{1}{4}(b\!-\!a)(1\!-\!z)^2\!-\!\beta^2\right]},
\end{align}
and
\begin{align}\label{eq.Iq2}
I_{q^2} & \equiv \frac{1}{2} \beta\left[\beta(\beta-1)K^2+M^2\right] f\left(\beta, \lambda,2\right) \nonumber \\
& =\frac{1}{2} \int \mathrm{d}a \mathrm{d}b \mathrm{d}v_1 \mathrm{d}u_1 \frac{\rho[z(a, b, \beta)]}{(b-a)\left(u_2 v_2\right)^{1 / 2}} \nonumber\\
&\times \frac{\lambda\beta^2[\beta(\beta-1)+\lambda] }{\left[\lambda\left(1\!-\!u_2\!-\!v_2\right)\!+\!a\!+\!\frac{1}{4}(b\!-\!a)(1\!-\!z)^2\!-\!\beta^2\right]^2}.
\end{align}
Finally, making $\beta=x$, the axial-vector diquark's chiral-odd PDA becomes 
\begin{equation}
n_{1^+}^o  \varphi^{odd}_{1^+}(x,\lambda)= I_{q^4}(x,\lambda,1) -I_{q^2}(x,\lambda,1).
\end{equation}

Figure~\ref{fig:AVDiquarkPDA} displays the leading-twist chiral-even and chiral-odd PDAs of the axial-vector diquark, together with the asymptotic QCD distribution. Both amplitudes are symmetric under $x \leftrightarrow 1-x$, implying that neither constituent quark is preferred on average. The chiral-even PDA is relatively close to the asymptotic form, exhibiting a pronounced maximum around $x=1/2$, which indicates a preference for configurations in which the momentum is distributed nearly equally between the quarks. In contrast, the chiral-odd PDA is considerably broader and displays enhanced support away from the central region, signaling stronger non-perturbative broadening in the tensor channel. This behaviour mirrors the pattern observed in vector mesons, where the chiral-odd PDA is also broader than its chiral-even counterpart~\cite{Chernyak:1983ej, Ball:1996tb}.

Within the present algebraic framework, both the chiral-even and chiral-odd axial-vector diquark PDAs depend on the diquark virtuality through the dimensionless ratio $\lambda=M^2/K^2$. At fixed dressed-quark mass $M$, varying $K^2$ changes the relative weight of the mass and momentum-dependent terms in the Feynman-parameter integrals, modifying the width of the resulting distributions. This dependence persists after normalization and therefore reflects a change in the longitudinal momentum-sharing profile, rather than merely an overall rescaling. The algebraic representation employed here thus retains a connection between the diquark virtuality and the shape of both PDAs.


\subsection{Proton PDA in the axial-vector diquark approximation}

\subsubsection{Chiral-even contribution}

First, for $\mathcal{L}_1^{1+}$, the Mellin moments of the nucleon are
\begin{align}
\label{MMs_algebraic_nucleon_1}
\frac{1}{2i} f_P &(n \cdot p)^2 \slashed{n} \expval{x_1^l x_2^m} = \int \frac{\mathrm{d}^4 K}{(2\pi)^4} \; \left( \frac{n\cdot p_1}{n \cdot P} \right)^l \gamma_\nu \nonumber \\
&
\times \Tr \left[ \gamma_\nu O^3_\varphi S(p_1) \mathcal{A}_\rho^j \right] \Delta_{\mu \rho}^1(K) D^{m\mu}_{1+} (K^2),
\end{align}
where $O^3_\varphi = \slashed{n} H_{\uparrow}$. The quark--axial-vector-diquark correlation amplitude is defined by
\begin{equation}
\mathcal{A}_\rho^j=\frac{1}{6} r_A o^j \gamma_5\left[\gamma_\rho-i r_P P_\rho \right] \int_{-1}^{1}\mathrm{d}z \rho(z)\Hat{\sigma}_{\Lambda_{1p}} (K_{+P}^2),
\end{equation}
where the isospin coefficients are
\begin{subequations}
\begin{align}
(I=1,\,I_z=1):\quad& o^+ = 2, \\
(I=1,\,I_z=0):\quad& o^0 = -1,
\end{align}
\end{subequations}
and $r_P = 13/87$. The parameter $r_A$ measures the relative $1^+:0^+$ diquark strengths in the Faddeev amplitude.

Moreover, the axial-vector diquark propagator is defined by
\begin{subequations}
\begin{align}
\Delta_{\mu \rho}^1 (K) =& T_{\mu \rho}(K)    \sigma_{M_1}(K^2), \\
T_{\mu \nu}(K) =& \delta_{\mu \nu}  -  \frac{K_\mu K_\nu}{K^2},\\
\sigma_M(s) =& \frac{1}{s+M^2}.
\end{align}
\end{subequations}

To compute Eq.~\eqref{MMs_algebraic_nucleon_1}, we first evaluate the trace
\begin{align}\label{trace_1_nucleon}
&\Tr \left[\gamma_\nu O^3_\varphi S(p_1) \mathcal{A}_\rho^j
\right]\Delta_{\mu \rho}^1 (K)=\frac{\Lambda_{1p}^2}{12}r_A o^j \int_{-1}^{1}\mathrm{d}z \rho(z) \nonumber\\
&\times \Tr \left[\gamma_\nu \slashed{n} (1\!+\!\gamma_5)(-i\gamma \!\cdot\! p_1 \!+\! M) \gamma_5(\gamma_\rho\! -\! i r_P P_\rho)\right]T_{\mu \rho} (K) \nonumber \\
&\times\sigma_{\Lambda_{1p}} (K_{+P}^2) \sigma_M(p_{1}^{2})
\sigma_{M_1} (K^2),
\end{align}
encountering a denominator of the form
\begin{align}\label{Eq.Denominator.ChiralEven-Nucleon}
&\sigma_{\Lambda_{1p}} (K_{+P}^2)
\sigma_M(p_{1}^{2})
\sigma_{M_1} (K^2)
=\nonumber\\
&\Gamma(3)
\int_0^1
\frac{\delta(1-a-b-c)\mathrm{d}a \mathrm{d}b \mathrm{d}c
}{\left[a(K_{+P}^2 + \Lambda^{2}_P)+b(p_{1}^{2} + M^2)+c(K^2 + M_1^2)
\right]^3}.
\end{align}

After integrating over the Dirac delta function and rearranging terms, the denominator becomes
\begin{align}
a(&K_{+P}^2 + \Lambda^{1\,2}_P) + b(p_{1}^{2} + M^2) + c(K^2 + M_1^2) \nonumber\\
&=
\left[ K - \left( b + \frac{1-z}{2} a \right) P \right]^2 \nonumber\\
&
+ P^2 \left[ b + \left( \frac{1-z}{2} \right)^2 a - \left( b + \frac{1-z}{2} a \right)^2 \right] \nonumber\\
&
+ a(\Lambda^{2}_P - M_1^2) + b(M^2-M_1^2) + M_1^2.
\end{align}

Defining $\eta= b + \frac{1-z}{2} a$, the denominator can be rewritten as
\begin{align}
a(&K_{+P}^2 + \Lambda^{1\,2}_P) + b(p_{1}^{2} + M^2) + c(K^2 + M_1^2) \nonumber\\
&=
\left( K - \eta P\right)^2 + P^2 \left[ \eta(1-\eta) - \frac14 (1-z^2)a \right] \nonumber\\
&
+ a(\Lambda^{2}_P - M_1^2) + b(M^2-M_1^2) + M_1^2 .
\end{align}

Substituting this expression into Eq.~\eqref{trace_1_nucleon}, and then inserting the result into Eq.~\eqref{MMs_algebraic_nucleon_1}, the Mellin moments become
\begin{align}\label{MMs_algebraic_nucleon_2}
\frac{1}{2i} & f_P (n \cdot p)^2 \slashed{n}\expval{x_1^l x_2^m}= \frac{1}{6} r_A o^j \nonumber \\
&
\times \int_{-1}^{1} \mathrm{d}z \rho(z) \Lambda^{2}_{P} \int_0^1 \mathrm{d}a \int_0^{1-a} \mathrm{d}b \int \frac{\mathrm{d}^4K}{(2\pi)^4}\left(\frac{n\!\cdot \! p_1}{n \! \cdot \! P} \right)^l \nonumber \\
&\times \frac{
\gamma_\nu \Tr \left[\gamma_\nu \slashed{n} (1+\gamma_5) (-i\gamma \cdot p_1 + M)  \gamma_5  (\gamma_\rho - i r_P P_\rho)
\right]}{ D_2^3} \nonumber \\
&\times T_{\mu \rho} (K) D_{1^+}^{m \mu}(K^2),
\end{align}
where the denominator is defined by
\begin{align}
D_2 &= \left( K - \eta P\right)^2+ P^2 \Big(\eta(1-\eta)-\frac14(1-z^2)a \Big)\Big. \nonumber \\
&
\Big.+a(\Lambda^{2}_P - M_1^2)+b(M^2-M_1^2)+M_1^2.
\end{align}

By substituting the Mellin moments calculated for the axial-vector diquark, Eq.~\eqref{MMs_algebraic_axialvectordiquark}, performing the shift $K \rightarrow K + \eta P$, carrying out the corresponding index contractions, integrating over $K$, and introducing suitable changes of variables and order of integration, the Mellin moments of the nucleon are obtained
\begin{align}\label{MMs_algebraic_nucleon_9}
n_{1^+}^{e'}f_P& (n \cdot p)^2  \expval{x_1^l x_2^m}= \nonumber \\
&\int_{0}^{1} \mathrm{d}\eta(1-\eta)^l \int_0^{\eta} \mathrm{d}\Tilde{\beta}\Tilde{\beta}^m \int_0^{\eta}  \mathrm{d}b\int_{\eta-b}^{1-b}\frac{\mathrm{d}a}{a\eta^3}\nonumber\\
&\times\frac{\Tilde{\beta}(\eta-\Tilde{\beta}) \rho\left[z(a,b,\eta)\right]\left[M r_P - 2(1-\eta)\right]}{ D_2'},
\end{align}
where the denominator is defined by
\begin{align}
\mathcal{D}_2 &= P^2\left[\eta(1-\eta)-\tfrac14(1-z^2)a\right] \nonumber \\
&
+a(\Lambda^{2}_P - M_1^2) +b(M^2-M_1^2)+M_1^2,
\end{align}
with $ z(a,b,\eta)=1 - 2\frac{\eta-b}{a}$. Comparing the above expression with the definition of the Mellin moments, one can identify the proton PDA
\begin{align}
\psi_{1^+}^{\mathrm{even}}&(1-\eta,\Tilde{\beta}) = \mathcal{N}_{1^+}^{\mathrm{e}} \int_0^{\eta}  \mathrm{d}b
\int_{\eta\!-\!b}^{1\!-\!b} \frac{\mathrm{d}a}{a\eta^3} \nonumber \\
&
\times \frac{
\Tilde{\beta}(\eta\!-\!\Tilde{\beta}) \rho\left[z(a,b,\eta)\right]
\left[Mr_P \!-\!2(1\!-\!\eta)\right]}{\mathcal{D}_{2}^2},
\end{align}
with normalization constant
\begin{equation}
\mathcal{N}_{1^+}^{\mathrm{even}} = \left[ \int_0^1 \mathrm{d}\eta \int_0^{1-\eta} \mathrm{d}\Tilde{\beta} \; \psi_{1^+}^{\mathrm{even}}(\eta,\Tilde{\beta}) \right]^{-1}.
\end{equation}

\subsubsection{Chiral-odd contribution}

The chiral-odd axial-vector diquark contribution to the proton PDA is obtained from the Mellin moments
\begin{align}
\label{Eq.MMs.ChiralOdd.Nucleon1}
\frac{1}{2 i} f_P & (n\!\cdot\! P)^2 \slashed{n}\, \langle x_1^l x_2^m \rangle
=
\int \frac{d^4K}{(2\pi)^4} \left(\frac{n\!\cdot\! p_2}{n\!\cdot\! P}\right)^l \nonumber \\
&
\times \gamma_\nu {\rm Tr} \left[ \gamma_\nu O_{T}^{3} S_u(p_2) A_\rho^j \right] \Delta_{\mu\rho}^{1}(K) D_{1^+O}^{m\mu}(K^2),
\end{align}
where $p_2=P-K$ and
$\Delta_{\mu\rho}^{1}(K)=T_{\mu\rho}(K)\sigma_{M_1}(K^2)$, with $T_{\mu\rho}(K)=\delta_{\mu\rho}-K_\mu K_\rho/K^2$. The quark--axial-vector-diquark correlation amplitude is modeled as
\begin{equation}
\label{Eq.Quark.Diquark.Amplitude}
A_\rho^j= \frac{nr_A}{6} \gamma_5 \left( \gamma_\rho\!-\!i r_p P_\rho \right) \int_{-1}^{1} dz \rho(z) \Lambda_{1p}^2 \sigma_{\Lambda_{1p}}(K_{+P}^2),
\end{equation}
whereas the chiral-odd Mellin moments of the axial-vector diquark are defined in Eq.~\eqref{eq6_avodd}. Upon substituting Eqs.~\eqref{Eq.Quark.Diquark.Amplitude} and \eqref{eq6_avodd} into Eq.~\eqref{Eq.MMs.ChiralOdd.Nucleon1}, the denominator can be written as
\begin{equation}
\sigma_{\Lambda_{1p}}(K_{+P}^2) \sigma(p_2^2) \sigma_{M_1}(K^2) = \int_0^1 dc \int_0^{1-c} de \frac{\Gamma(3)}{D_{3}^2},
\end{equation}
with
\begin{align}
D_{3} &= (K-\eta P)^2+P^2\left[\eta(1-\eta)-\frac14(1-z^2)c\right] \nonumber \\
&
+c(\Lambda_{1p}^2-M_1^2)+e(M^2-M_1^2)+M_1^2.
\end{align}

To evaluate the integral over the $K$ momentum, we first perform the shift $K \rightarrow K + \eta P$. After substituting the explicit expressions for the quark--diquark amplitude and the axial-vector diquark Mellin moments, carrying out the corresponding index contractions and Dirac traces, and performing the shift $K \rightarrow K + \eta P$, the numerator of Eq.~\eqref{Eq.MMs.ChiralOdd.Nucleon1} reduces to
\begin{align}
&\left[\frac{n\!\cdot\!(P\!-\!K)} {n\!\cdot\!P}\right]^l\frac{(n\!\cdot\!K)^m}{(n\!\cdot\!P)^{m+1}}\gamma_\nu{\rm Tr}[\gamma_\nu
O_{T}^{3}S_u(p_2)A_\rho^j]T_{\mu\rho}(K) \nonumber\\
&\times \left(\delta_{\mu\beta}n\!\cdot\!K-n_\mu K_\beta\right)\delta_{\beta\nu}^{\perp}=\nonumber\\
&-2(1\!-\!\eta)^l\eta^m\slashed n\Big[r_p K^2\!+\!r_p(1-2\eta)\eta P^2\!+\!2\eta M^2\Big].
\end{align}

Consequently, the Mellin moments reduce to
\begin{align}
f_P&(n\!\cdot\!P)^2 \langle x_1^l x_2^m\rangle \sim \int_{-1}^{1}dz \rho(z) (1-\eta)^l \eta^m \int_0^1 d\beta \beta^m \int_0^1 dc \nonumber \\
&
\times \int_0^{1-c}de \Big[ I_{q^4}(\beta,\lambda) \!-\! I_{q^2}(\beta,\lambda) \Big] \nonumber \\
&
\times \frac{r_p(1\!-\!2\eta)\eta P^2\!+\!2\eta M^2}{\mathcal{D'}_{3}^{2}},
\end{align}
where the denominator becomes
\begin{align}
\mathcal{D'}_{3} &= P^2\left[\eta(1-\eta)-\frac14(1-z^2)c\right] \nonumber \\
&
+c(\Lambda_{1p}^2-M_1^2)+e(M^2-M_1^2)+M_1^2.
\end{align}
Introducing the variables $\widetilde\beta=\eta\beta$ and $z=1-\frac{\eta-e}{c}$ rearranging the integration domain,
and assuming the canonical form of the Mellin moments, the chiral-odd axial-vector contribution to the proton PDA is identified as
\begin{align}
\psi_{1^+}^{\rm odd}&(1-\eta,\widetilde\beta)= \mathcal N_{1^+}^{\rm o} \int_0^\eta dc \int_0^{1-c}de\, \rho[z(c,e,\eta)]\nonumber \\
&
\times \Big[ I_{q^4}(\beta,\lambda) - I_{q^2}(\beta,\lambda)
\Big] \frac{r_p(1-2\eta)P^2+2M^2}{\mathcal{D'}_{3}^2},
\end{align}
where $\mathcal N_{1^+}^{\rm o}$ is fixed by the normalization condition of the proton PDA.

\begin{figure}[!t]
\centering
\includegraphics[width=0.39\textwidth]{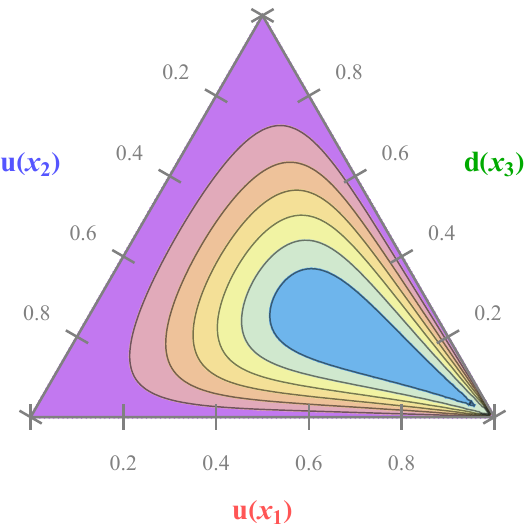} \\
\includegraphics[width=0.39\textwidth]{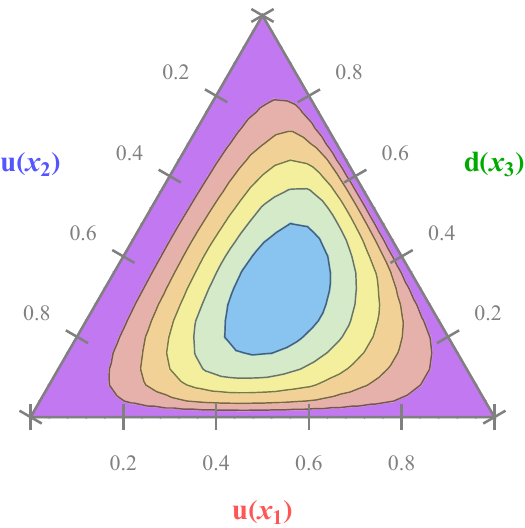} \\
\includegraphics[width=0.39\textwidth]{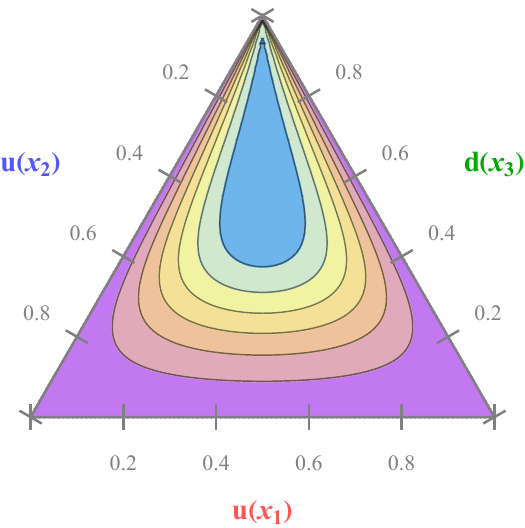}  \\
\caption{\label{fig:AVDiquarkContributions} Axial-vector diquark distribution amplitudes. From top to bottom: $\{ud\}$ chiral-even, $\{ud\}$ chiral-odd, and $\{uu\}$ chiral-even. The distribution is displayed in barycentric coordinates $(x_1,x_2,x_3)$, with $x_1+x_2+x_3=1$.}
\end{figure}

The resulting distributions associated with the three possible axial-vector diquark configurations are displayed in Fig.~\ref{fig:AVDiquarkContributions}. As expected from the structure of the underlying axial-vector diquark PDAs, the chiral-even and chiral-odd channels produce markedly different momentum distributions. In particular, the chiral-odd component generates a broader and less asymmetric distribution, whereas the chiral-even contribution remains more strongly localized. The interplay between these two structures enhances the breaking of the symmetry under the interchange $x_2 \leftrightarrow x_3$, leading to a redistribution of longitudinal momentum among the valence quarks.

Having determined the chiral-even and chiral-odd quark--axial-vector-diquark contributions, we can now construct the proton's PDA within the so-called axial-vector diquark approximation. We denote by $\varphi_{AV}^{i}$ the normalized contribution from the axial-vector diquark configuration in which quark $i$ acts as the spectator. The relative weights of the three spectator configurations follow from the spin--isospin structure of the nucleon Faddeev amplitude. The total axial-vector contribution is therefore given by
\begin{equation}
\label{eq:SpinIsospinWeights}
\varphi_{AV} = \frac{1}{2} \left( -\varphi_{AV}^{1} +\varphi_{AV}^{2} +2\varphi_{AV}^{3} \right).
\end{equation}
Equation~\eqref{eq:SpinIsospinWeights} preserves the permutation symmetry of the proton wave function while incorporating the spin--isospin weights associated with the different axial-vector-diquark configurations.

\begin{figure}[!t]
\centering
\includegraphics[width=0.39\textwidth]{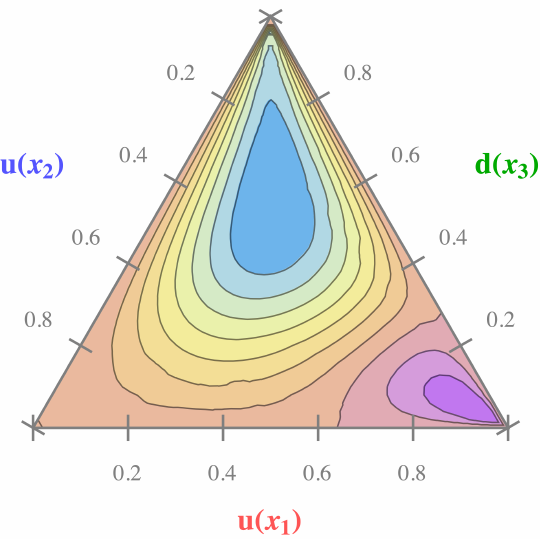} \\
\caption{\label{fig:NucleonPDAAVOnly} Leading-twist proton PDA obtained in the axial-vector-diquark approximation, \emph{i.e.} retaining only axial-vector correlations in the nucleon Faddeev wave function. Compared with the scalar-diquark result shown in Fig.~\ref{fig:NucleonPDAScalarOnly}, the distribution is broader and less asymmetric, reflecting the redistribution of longitudinal momentum induced by the spin--1 diquark sector. Besides, the distribution is not positive definite, exhibiting localized negative regions generated by destructive interference among the different axial-vector spectator contributions. The distribution is displayed in barycentric coordinates $(x_1,x_2,x_3)$, with $x_1+x_2+x_3=1$.}
\end{figure}

Figure~\ref{fig:NucleonPDAAVOnly} displays the proton PDA obtained by retaining only axial-vector diquark correlations in the nucleon Faddeev amplitude. In contrast to the scalar-diquark approximation shown in Fig.~\ref{fig:NucleonPDAScalarOnly}, the resulting distribution is less asymmetric and exhibits a broader support over the kinematically allowed region. Besides, the distribution is not positive definite, exhibiting localized regions where it changes sign. Since the PDA is a quantum-mechanical amplitude rather than a probability density, these negative values signal destructive interference among the different axial-vector spectator contributions entering Eq.~\eqref{eq:SpinIsospinWeights}. They therefore reflect the non-trivial spin--flavour structure of the axial-vector sector and illustrate how the various spectator configurations combine to redistribute longitudinal momentum within the proton.

\section{Proton PDA with scalar and axial-vector diquarks}
\label{sec:PDAtotal}

Having determined the scalar-diquark component together with the chiral-even and chiral-odd axial-vector contributions, we can now construct the complete proton PDA. The relative strength of the axial-vector sector is controlled by the parameter $r_A$ introduced in Eq.~\eqref{eq:AmuDef}. Following continuum Faddeev-equation studies, $r_A$ is chosen such that scalar-diquark correlations account for approximately 65\% of the nucleon wave-function norm,
\begin{align}
\label{eq:RaFixingCondition} 
\int \mathcal{D}x \,
\varphi_S^2(x_1,x_2,x_3) &= (0.65 \pm 0.05) \nonumber \\
&
\times \int \mathcal{D}x \, \varphi^2(x_1,x_2,x_3).
\end{align}
Implementing Eq.~\eqref{eq:RaFixingCondition} leads to
\begin{equation}
r_A = 0.30 \pm 0.03.
\end{equation}

\begin{figure}[!t]
\centering
\includegraphics[width=0.4\textwidth]{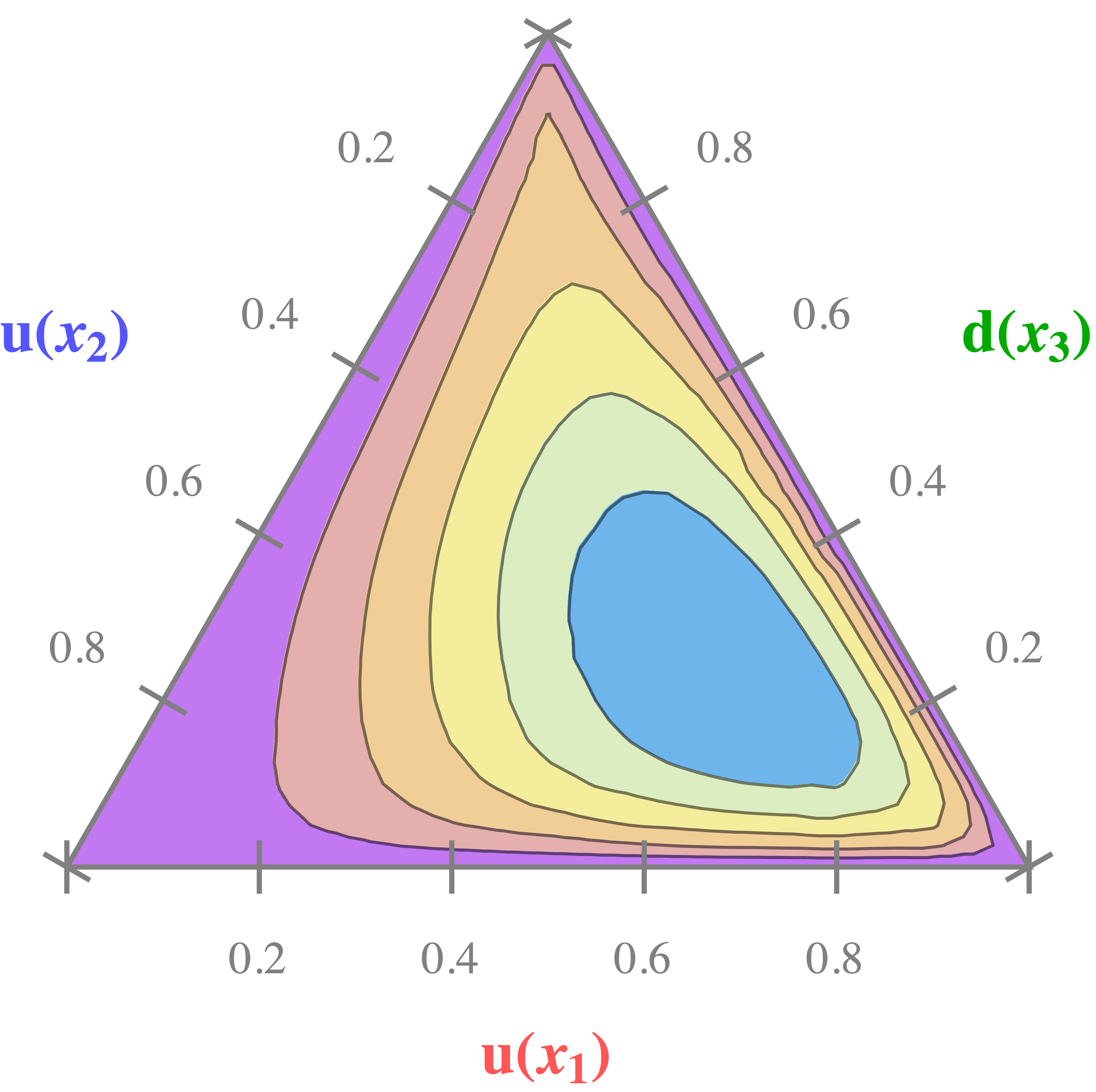}
\caption{\label{fig:TotalNucleonDA} Leading-twist nucleon parton distribution amplitude obtained from the complete quark--diquark Faddeev wave function, including both scalar ($0^+$) and axial-vector ($1^+$) diquark correlations. The distribution is displayed in barycentric coordinates $(x_1,x_2,x_3)$, with $x_1+x_2+x_3=1$.}
\end{figure}

We first normalize the scalar- and axial-vector-diquark contributions independently,
\begin{equation}
    \int[dx]\,\varphi_S(x_i)=
\int[dx]\,\varphi_{AV}(x_i)=1.
\end{equation}
Motivated by continuum Faddeev-equation studies, we then construct
the complete proton PDA as
\begin{equation}
    \varphi(x_i)=0.65\,\varphi_S(x_i)+0.35\,\varphi_{AV}(x_i).
\end{equation}
The resulting distribution is shown in Fig.~\ref{fig:TotalNucleonDA}. Compared with the scalar-diquark result in Fig.~\ref{fig:NucleonPDAScalarOnly}, the inclusion of axial-vector correlations broadens the distribution and redistributes longitudinal momentum among the three valence quarks.

The scalar-diquark contribution alone produces a strongly asymmetric PDA, with the maximum located in a region where the bystander quark carries approximately one half of the proton longitudinal momentum. This is reflected in the first Mellin moments,
\begin{equation}
\langle x_1\rangle = 0.500,\quad
\langle x_2\rangle = 0.250,\quad
\langle x_3\rangle = 0.250.
\end{equation}
This pattern is consistent with a compact scalar diquark carrying approximately half of the proton momentum, while the remaining momentum is carried by the bystander quark.

Axial-vector correlations significantly modify this picture. In the complete calculation, the first Mellin moments become
\begin{equation}
\langle x_1\rangle = 0.404,\quad
\langle x_2\rangle = 0.266,\quad
\langle x_3\rangle = 0.331.
\end{equation}
Thus, the axial-vector sector reduces the strong spectator-quark dominance generated by scalar diquarks and shifts strength into alternative spin-flavour configurations. This effect is also visible in Figs.~\ref{fig:NucleonPDAScalarOnly},~\ref{fig:AVDiquarkContributions} and~\ref{fig:TotalNucleonDA}, where the complete PDA exhibits broader support than the scalar-diquark and axial-vector contributions separately.

The maximum of the full proton PDA is located at
\begin{equation}
(x_1,x_2,x_3) \simeq (0.55,0.20,0.25),
\end{equation}
which remains substantially displaced from the asymptotic maximum at $(1/3,1/3,1/3)$. Hence, even after axial-vector correlations are included, the distribution retains a strong imprint of non-perturbative quark--diquark dynamics.

Table~\ref{tab:MellinMomentComparison} compares the first Mellin moments obtained here with lattice-QCD calculations and the PTIR-based quark--diquark analysis of Ref.~\cite{Mezrag:2017znp}. The present calculation reproduces the qualitative asymmetry found in these approaches, with the proton PDA shifted toward the large-$x_1$ region. Quantitatively, our unevolved result yields a larger value of $\langle x_1\rangle$ and a smaller value of $\langle x_2\rangle$ than the quoted estimates, whereas $\langle x_3\rangle$ is somewhat
larger and remains close to the symmetric value $1/3$. However, the model results listed in Table~\ref{tab:MellinMomentComparison} correspond to the hadronic scale, $\mu_0=0.51\,\mathrm{GeV}$, and therefore should not be compared quantitatively with determinations reported at higher renormalization scales. A scale-matched comparison is presented in the following section, where the complete proton PDA
is evolved to $\mu=2\,\mathrm{GeV}$. Evolution towards higher scales is expected to soften this pattern, driving the first moments towards their conformal-limit values, $\langle x_i\rangle=1/3$. Consequently, the observed differences cannot be attributed solely to the underlying quark--diquark dynamics; a quantitative comparison requires evolving the results to a common scale.


Nevertheless, the inclusion of axial-vector correlations moves the first moments significantly away from the scalar-diquark limit and toward the values obtained in other non-perturbative approaches. This demonstrates that axial-vector diquarks are not merely a quantitative correction: they provide an essential mechanism for redistributing longitudinal momentum and spin within the proton. Overall, scalar correlations generate the dominant asymmetry of the proton PDA, whereas axial-vector correlations broaden the distribution and reduce the momentum imbalance among the valence quarks.

\begin{table}[!t]
\centering
\begin{tabular}{|l|c|c|c|}
\hline
 & $\langle x_1\rangle_u$ & $\langle x_2\rangle_u$ & $\langle x_3\rangle_d$ \\
\hline
Asymptotic & $1/3$ & $1/3$ & $1/3$ \\
\hline
LQCD~\cite{RQCD:2019hps}
& $0.396^{+0.007}_{-0.006}$
& $0.311(5)$
& $0.293^{+0.005}_{-0.006}$ \\

LQCD~\cite{Bali:2024oxg}
& $0.400^{+0.009}_{-0.008}$
& $0.309(5)$
& $0.290^{+0.006}_{-0.007}$ \\
\hline
Ref.~\cite{Mezrag:2017znp}, proton & $0.392(5)$ & $0.291(2)$ & $0.316(4)$ \\
Ref.~\cite{Mezrag:2017znp}, scalar only & $0.434$ & $0.283$ & $0.283$ \\
\hline
This work, proton & $0.404$ & $0.266$ & $0.331$ \\
This work, scalar only & $0.450$ & $0.250$ & $0.250$ \\
\hline
\end{tabular}
\caption{\label{tab:MellinMomentComparison} First Mellin moments of the proton leading-twist PDA. Results from the present work are compared with the asymptotic prediction, lattice-QCD calculations, and the PTIR-based quark--diquark analysis of Ref.~\cite{Mezrag:2017znp}.}
\end{table}


\section{QCD evolution of the proton distribution amplitude}
\label{sec:pda-evolution}

The proton PDA obtained from the quark--diquark calculation is defined at the hadronic scale $\mu_0=0.51\,\mathrm{GeV}$.  In order to compare it with results quoted at a conventional perturbative scale, we evolve its normalized shape to $\mu=2\,\mathrm{GeV}$ at leading order (LO). The scale dependence of baryon distribution amplitudes is governed by the Brodsky--Lepage evolution equation~\cite{Lepage:1980fj}. Rather than evolving the values of the PDA point by point, we first project it onto a polynomial basis adapted to the one-loop evolution kernel of the three-quark light-ray operator.  The projection decomposes the non-perturbative distribution into modes with a definite total polynomial degree.  At fixed degree, the modes that mix under renormalization are subsequently diagonalized, so that the resulting eigenmodes evolve multiplicatively.

We use the momentum-fraction convention introduced above, with $x_1+x_2+x_3=1$, and the integration measure
\begin{equation}
  \int[\mathrm{d}x]
  \equiv
  \int_0^1\!\mathrm{d}x_1
  \int_0^1\!\mathrm{d}x_2
  \int_0^1\!\mathrm{d}x_3\,
  \delta(1-x_1-x_2-x_3).
  \label{eq:evol-measure}
\end{equation}
For the normalized PDA, $\int[\mathrm{d}x]\,\varphi(x_i;\mu)=1$, the asymptotic weight is
\begin{equation}
  w(x_i)=120x_1x_2x_3,
  \qquad
  \int[\mathrm{d}x],w(x_i)=1.
  \label{eq:evol-weight}
\end{equation}
Following Ref.~\cite{Braun:1999te}, the polynomial basis is chosen as
\begin{align}
  P_{Nn}(x_i)
  ={}&(N+n+4)(x_2+x_3)^n
  P_{N-n}^{(2n+3,1)}(2x_1-1)
  \nonumber\\
  &\times
  C_n^{3/2}\!\left(\frac{x_3-x_2}{x_2+x_3}\right),
  \qquad 0\leq n\leq N,
  \label{eq:evol-basis}
\end{align}
where $P_k^{(a,b)}$ and $C_n^{3/2}$ denote Jacobi and Gegenbauer polynomials, respectively.  These polynomials are orthogonal with respect to $w(x_i)$; their squared norms are
\begin{equation}
  \mathcal{N}_{Nn}
  =
  \int[\mathrm{d}x],w(x_i)P_{Nn}^{,2}(x_i).
  \label{eq:evol-norm}
\end{equation}
Our choice singles out $x_1$ as the momentum fraction of the bystander
quark, while $(x_2,x_3)$ describe the two quarks forming the correlated
diquark. Accordingly, the Jacobi polynomial describes the momentum
partition between the bystander quark and the diquark, whereas the
Gegenbauer polynomial resolves the internal momentum sharing within
the diquark. The sign convention in the Gegenbauer argument only
changes the sign of the odd-$n$ basis elements and has no physical
consequence when used consistently throughout the projection,
evolution, and reconstruction.

Accordingly, the numerical PDA at the initial scale is projected through
\begin{equation}
  B_{Nn}(\mu_0)
  =
  \int[\mathrm{d}x],P_{Nn}(x_i)\varphi(x_i;\mu_0).
  \label{eq:evol-projection}
\end{equation}
The index $N$ labels the total polynomial degree.  Hence the evolution kernel does not couple sectors with different $N$, although the $N+1$ moments within the same sector generally mix.  Writing
\begin{equation}
  \boldsymbol{B}_N
  =
  (B_{N0},B_{N1},\ldots,B_{NN})^{\mathrm T},
\end{equation}
we diagonalize each fixed-$N$ block and define the eigenmoments
\begin{equation}
  \boldsymbol{E}_N(\mu)
  =
  \boldsymbol{V}_N\boldsymbol{B}_N(\mu).
  \label{eq:evol-diagonalization}
\end{equation}
Here $\boldsymbol{V}_N$ is the change-of-basis matrix obtained from the LO three-quark evolution kernel.  The components $E_{Nk}$ are multiplicatively renormalizable and therefore evolve independently.

For $n_f=4$ active flavours, we use
\begin{equation}
  \beta_0=11-\frac{2}{3}n_f=\frac{25}{3},
  \qquad
  \Lambda_{\mathrm{QCD}}=0.234\,\mathrm{GeV},
  \label{eq:evol-parameters}
\end{equation}
and the one-loop running coupling.  Since we evolve the normalized shape of the PDA, rather than its dimensionful normalization constant, the LO evolution of every eigenmoment reads
\begin{equation}
  E_{Nk}(\mu)
  =
  E_{Nk}(\mu_0)
  \left[
    \frac{\ln(\mu^2/\Lambda_{\mathrm{QCD}}^2)}
         {\ln(\mu_0^2/\Lambda_{\mathrm{QCD}}^2)}
  \right]^{-[\gamma_{Nk}-\gamma_{00}]/\beta_0},
  \label{eq:eigenmoment-evolution}
\end{equation}
with $\gamma_{00}=\frac{2}{3}$. The subtraction of $\gamma_{00}$ removes the common scale dependence carried by the overall three-quark normalization.  For the scales employed here, the logarithmic ratio in Eq.~\eqref{eq:eigenmoment-evolution} is
\begin{equation}
  \frac{\ln(\mu^2/\Lambda_{\mathrm{QCD}}^2)}
       {\ln(\mu_0^2/\Lambda_{\mathrm{QCD}}^2)}
  =2.75396.
\end{equation}
After evolving the eigenmoments, we return to the original polynomial moments using
\begin{equation}
  \boldsymbol{B}_N(\mu)
  =
  \boldsymbol{V}_N^{-1}\boldsymbol{E}_N(\mu),
\end{equation}
and reconstruct the PDA as
\begin{equation}
  \varphi^{(N_{\max})}(x_i;\mu)
  =
  120x_1x_2x_3
  \left[
    1+
    \sum_{N=1}^{N_{\max}}
    \sum_{n=0}^{N}
    \frac{B_{Nn}(\mu)}{\mathcal{N}_{Nn}}P_{Nn}(x_i)
  \right].
  \label{eq:evolved-reconstruction}
\end{equation}
We retain all modes through $N_{\max}=5$.  At the initial scale, the degree-five reconstruction yields an integrated absolute error of $0.0396$ and a relative $L^2$ error of $0.0444$ with respect to the numerical grid.  Moreover, the inclusion of degree five changes the evolved distribution by only about $2.4\%$ relative to the degree-four result. We therefore use $N_{\max}=5$ as our central truncation and regard the comparison with lower orders as a numerical stability test.

The resulting first Mellin moments are collected in Table~\ref{tab:evolved-first-moments}. The evolution decreases the momentum fraction carried by the first quark and increases that carried by the second, whereas the third moment remains close to one third.  Thus, the initial hierarchy is reduced and the PDA moves towards the symmetric conformal limit, although sizeable non-perturbative asymmetry remains at $2\,\mathrm{GeV}$. This is expected because perturbative evolution is logarithmically slow.

\begin{table}[t]
  \centering
  \caption{First Mellin moments of the normalized proton PDA at the hadronic
  scale and after LO evolution to $2\,\mathrm{GeV}$.  The last column gives
  the change produced by the evolution.}
  \label{tab:evolved-first-moments}
  \begin{tabular}{lccc}
    \hline\hline
    Moment & $\mu_0=0.51\,\mathrm{GeV}$ & $\mu=2\,\mathrm{GeV}$
           & Difference \\
    \hline
    $\langle x_1\rangle$ & 0.403678 & 0.385664 & $-0.018014$ \\
    $\langle x_2\rangle$ & 0.265532 & 0.284305 & $+0.018772$ \\
    $\langle x_3\rangle$ & 0.330790 & 0.330031 & $-0.000759$ \\
    \hline\hline
  \end{tabular}
\end{table}

\begin{figure}[t]
\centering
\includegraphics[width=0.4\textwidth]{PDA_NUCLEON_TOTAL.png}
\includegraphics[width=0.4\textwidth]{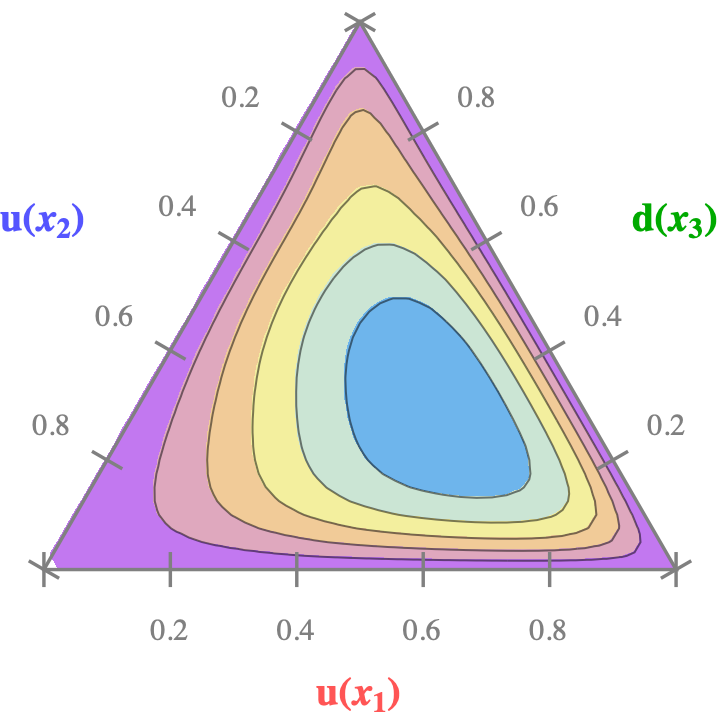}
\caption{Normalized proton PDA at the hadronic scale $\mu_0=0.51\,\mathrm{GeV}$ (top) and after LO evolution to
  $\mu=2\,\mathrm{GeV}$ (bottom).}
  \label{fig:pda-evolution}
\end{figure}

Figure~\ref{fig:pda-evolution} displays the degree-five reconstruction after evolution. In agreement with the behaviour of the first moments, the evolved PDA is smoother and less asymmetric, and its dominant support is shifted towards more balanced momentum sharing.  Nevertheless, it remains distinct from the asymptotic form
\begin{equation}
  \varphi_{\mathrm{asy}}(x_i)=120x_1x_2x_3,
\end{equation}
showing that the evolved distribution retains information about the non-perturbative quark--diquark dynamics at the hadronic scale.

\section{Summary}
\label{sec:summary}

We have presented an algebraic study of the leading-twist proton parton distribution amplitude within a Poincaré-covariant quark--diquark framework. The construction is based on perturbation-theory integral representations (PTIRs) for the quark, diquark, and quark--diquark correlation amplitudes, which enable the analytic evaluation of Mellin moments and the explicit reconstruction of the corresponding distribution amplitudes.

The analysis began with the scalar-diquark sector, for which compact analytical expressions were obtained for both the diquark PDA and its contribution to the proton distribution amplitude. This approximation produces a proton PDA strongly skewed toward configurations in which the spectator quark carries the largest longitudinal momentum fraction. Such a pattern is consistent with previous continuum studies of baryon structure based on covariant Faddeev equations and quark--diquark models, and reflects
the strongly correlated nature of the remaining quark pair.

We subsequently incorporated axial-vector diquark correlations and showed that the spin--1 sector naturally separates into chiral-even and chiral-odd
components. The corresponding axial-vector-diquark PDAs were derived analytically and included in the construction of the complete proton PDA. The scalar and axial-vector contributions were combined with relative weights of $65\%$ and $35\%$, respectively, after independently normalizing both components. The resulting proton PDA remains significantly skewed with respect to the asymptotic distribution,
\begin{equation}
    \varphi_{\mathrm{asy}}(x_1,x_2,x_3)
    =
    120\,x_1x_2x_3,
\end{equation}
but exhibits a broader and less asymmetric momentum distribution than that obtained in the scalar-diquark approximation alone.

The inclusion of axial-vector correlations redistributes longitudinal momentum among the valence quarks and thereby partially reduces the strong asymmetry generated by the scalar-diquark sector. This effect is reflected in both the shape of the proton PDA and its Mellin moments. At the hadronic scale $\mu_0=0.51\,\mathrm{GeV}$, the distribution possesses a maximum near
\begin{equation}
    (x_1,x_2,x_3)
    \simeq
    (0.55,0.20,0.25),
\end{equation}
which is displaced from the asymptotic maximum at $(1/3,1/3,1/3)$ and therefore highlights the persistence of strong non-perturbative correlations. The corresponding first Mellin moments are
\begin{equation}
    \big(
    \langle x_1\rangle,
    \langle x_2\rangle,
    \langle x_3\rangle
    \big)_{\mu_0}
    =
    (0.404,\,0.266,\,0.331).
\end{equation}
Hence, the first quark carries the largest average longitudinal momentum fraction, whereas the third moment is already close to the symmetric value
$1/3$.

In order to compare the model prediction with results quoted at conventional renormalization scales, we performed the leading-order QCD evolution of the normalized proton PDA from $\mu_0=0.51\,\mathrm{GeV}$ to $\mu=2\,\mathrm{GeV}$. The numerical distribution was projected onto an orthogonal Jacobi--Gegenbauer polynomial basis adapted to the one-loop evolution kernel of the three-quark light-ray operator. At each total polynomial degree, the corresponding mixing matrix was diagonalized, the resulting eigenmoments were evolved multiplicatively, and the evolved PDA was reconstructed after transforming back to the original polynomial basis. The expansion was truncated at $N_{\max}=5$, for which the reconstruction of the initial numerical PDA is stable and the inclusion of the highest retained degree changes the evolved distribution only at the few-percent level.

The evolution was implemented with $n_f=4$ active flavours and $\Lambda_{\mathrm{QCD}}=0.234\,\mathrm{GeV}$. At $\mu=2\,\mathrm{GeV}$, the first Mellin moments become
\begin{equation}
    \big(
    \langle x_1\rangle,
    \langle x_2\rangle,
    \langle x_3\rangle
    \big)_{2\,\mathrm{GeV}}
    =
    (0.386,\,0.284,\,0.330).
\end{equation}
Thus, evolution decreases $\langle x_1\rangle$, increases $\langle x_2\rangle$, and leaves $\langle x_3\rangle$ close to one third. The imbalance between the first two momentum fractions is consequently reduced, and the dominant support of the PDA moves toward configurations with more balanced momentum sharing. This behaviour is consistent with the suppression of non-asymptotic eigen-modes under QCD evolution.

The evolved distribution nevertheless remains asymmetric at $2\,\mathrm{GeV}$. This is expected because the scale dependence is logarithmic and the evolution interval is finite. Consequently, the PDA retains a clear imprint of the non-perturbative quark--diquark correlations present at the hadronic scale, while moving systematically toward the conformal limit. In particular, the evolution reduces the discrepancy between the model moments and those obtained in continuum and lattice-QCD
calculations quoted at $\mu=2\,\mathrm{GeV}$ \cite{Mezrag:2017znp,Braun:2014wpa,Bali:2015ykx}, thereby allowing a more meaningful comparison at a common renormalization scale.

Overall, the present calculation provides qualitative agreement with continuum and lattice-QCD studies: the proton PDA is shifted toward the large-$x_1$ region, and QCD evolution progressively softens this asymmetry. Some quantitative differences remain, particularly in the momentum redistribution between the first and second quarks. These differences may reflect the restricted momentum dependence and Dirac structure of the algebraic Faddeev amplitude, as well as contributions omitted from the
scalar- and axial-vector-diquark sectors. Within the retained truncation, scalar-diquark correlations generate the dominant asymmetry, axial-vector correlations partially redistribute the longitudinal momentum, and perturbative evolution further suppresses the non-asymptotic components.

Future extensions of this framework could incorporate more complete momentum-dependent Faddeev amplitudes and additional Dirac structures, allowing us to assess the effects of the components omitted in the present truncation and to extend the analysis to excited baryons and other members of the baryon spectrum. The evolved proton PDA also provides a suitable starting point for the calculation of hard exclusive observables. In particular, it may be combined with the perturbative hard-scattering kernel to study the proton electromagnetic Dirac form factor, $F_1(Q^2)$, at large momentum transfer, following the framework illustrated in Ref.~\cite{Rodriguez-Quintero:2018wma}. Evolving the PDA to the appropriate factorization scale would then make it possible to investigate how scalar- and axial-vector-diquark correlations influence an experimentally accessible
observable.

\begin{acknowledgments}
We are grateful to Craig D. Roberts for constructive discussions during the development of this work. We also acknowledge valuable discussions with Lei Chang, Minghui Ding, Si-xue Qin, and Jos\'e Rodr\'iguez-Quintero. B.A.Z. acknowledges SECIHTI (CVU No.~935777) for a postdoctoral fellowship through the program \textit{Posdoctorados por M\'exico}. C.M. acknowledges partial support from l'Agence Nationale de la Recherche (ANR) under Project No.~ANR-23-CE31-0019. J.S. acknowledges financial support from the Ministerio Espa\~nol de Ciencia e Innovaci\'on under Grant No.~PID2022-140440NB-C22, from the Junta de Andaluc\'ia under Contract No.~FQM-370, and from the PCI+D+i project ``Tecnolog\'ias avanzadas para la exploraci\'on del universo y sus componentes'' (Code AST22-0001). For the purpose of Open Access, the author has applied a CC-BY public copyright licence to any Author Accepted Manuscript (AAM) version
arising from this submission.
\end{acknowledgments}

\appendix
\section{Light-front kinematics and conventions}
\label{app:LF}

In the light-front formalism, it is convenient to introduce two auxiliary light-like four-vectors, $n^\mu$ and $p^\mu$, satisfying
\begin{equation}
n^2 = 0, \qquad p^2 = 0,
\end{equation}
with $p \cdot n \neq 0$ \cite{Brodsky:1997de, Diehl:2003ny}. 
The vector $n^\mu$ defines the light-front direction and is used to project the longitudinal components of four-momenta, while $p^\mu$ will be constructed such that it is also light-like and shares the same longitudinal projection as the hadron momentum.

Starting from the on-shell hadron momentum $P^\mu$, which satisfies
\begin{equation}
P^2 = M_N^2,
\end{equation}
we define a new vector $p^\mu$ as a linear combination of $P^\mu$ and $n^\mu$,
\begin{equation}
p^\mu = P^\mu - \alpha\, n^\mu,
\end{equation}
where $\alpha$ is a scalar coefficient to be determined. 
This form is the most general shift along the light-like direction $n^\mu$.

The value of $\alpha$ is fixed by requiring that $p^\mu$ becomes light-like, i.e., $p^2 = 0$. Explicitly,
\begin{equation}
p^2 = (P - \alpha n)^2 = P^2 - 2\alpha\, P\cdot n + \alpha^2 n^2.
\end{equation}
Since $n^\mu$ is light-like, $n^2 = 0$, the last term vanishes and one obtains
\begin{equation}
p^2 = M_N^2 - 2\alpha\, P\cdot n.
\end{equation}
Imposing the condition $p^2 = 0$, one immediately finds
\begin{equation}
\alpha = \frac{M_N^2}{2\,P\cdot n},
\end{equation}
so that
\begin{equation}
p^\mu = P^\mu - n^\mu \frac{M_N^2}{2\,P\cdot n}.
\end{equation}
Thus, $p^\mu$ is obtained from $P^\mu$ by removing the component responsible for its nonzero invariant mass, rendering it light-like.

Contracting this expression with $n^\mu$, one obtains
\begin{equation}
p \cdot n = P \cdot n - n^2 \frac{M_N^2}{2\,P\cdot n}.
\end{equation}
Using again $n^2 = 0$, it follows that
\begin{equation}
p \cdot n = P \cdot n,
\end{equation}
which shows that this construction preserves the longitudinal projection along the light-front direction. This property is essential, since $P \cdot n$ sets the scale of longitudinal momentum in the light-front framework.

With these vectors, any four-vector $k^\mu$ can be decomposed as
\begin{equation}
k^\mu = \frac{k \cdot n}{p\cdot n}\, p^\mu + \frac{k \cdot p}{p \cdot n}\, n^\mu + k_\perp^\mu,
\end{equation}
where the transverse component $k_\perp^\mu$ is defined through the orthogonality conditions
\begin{equation}
k_\perp \cdot n = k_\perp \cdot p = 0.
\end{equation}
This decomposition separates longitudinal and transverse degrees of freedom and is the basis of light-front kinematics \cite{Brodsky:1997de}.

\section{Complex-contour treatment of the scalar-diquark loop integral}
\label{app:diquarkContour}

We describe the contour prescription underlying the change of variables $k'=k-\beta K$ in the scalar-diquark calculation. In the Euclidean
nucleon rest frame,

\begin{equation}
P=(\boldsymbol{0},iM_N),
\quad
K=\frac23P-\ell,
\quad
\ell\in\mathbb{R}^4.
\end{equation}

The symmetric relative momentum $q=k-K/2$ is integrated over $\mathbb{R}^4$. Hence,
\begin{equation}
k'=q+\left(\frac12-\beta\right)K,
\end{equation}
and its temporal component follows the contour
\begin{equation}
k'_4=\bar{k}_4+i(1-2\beta)\frac{M_N}{3},
\quad
\bar{k}_4=q_4-\left(\frac12-\beta\right)\ell_4
\in\mathbb{R}.
\end{equation}

Thus, the change of variables does not immediately yield an integral over real Euclidean $k'$. To examine the singularities, define
\begin{equation}
a=\frac{1-z}{2},
\quad
\beta=y+av,
\quad
A=y+a^2v-\beta^2.
\end{equation}

The combined denominator is
\begin{equation}
D=k'^2+\mathcal{M}^2,
\quad
\mathcal{M}^2
=A K^2+M^2(u+y)+v\Lambda_\Gamma^2,
\end{equation}
with $u,v,y\geq0$, $u+v+y=1$, and $|z|\leq1$. The coefficient $A$ can be written as
\begin{equation}
A=uy+uv\,a^2+yv(1-a)^2\geq0,
\end{equation}
and satisfies
\begin{equation}
A+\left(\frac12-\beta\right)^2=\frac{u+y+vz^2}{4}.
\end{equation}

Consider a continuous deformation of the temporal contour to the real axis,
\begin{equation}
k'_4=t+ih,
\quad
t\in\mathbb{R},
\quad
h=s(1-2\beta)\frac{M_N}{3},
\quad
0\leq s\leq1.
\end{equation}
The spatial components of $k'$ remain real. Using
$\operatorname{Re}K^2=\ell^2-4M_N^2/9$, the real part of the denominator becomes
\begin{align}
\operatorname{Re}D
={}&\boldsymbol{k}'^{\,2}+t^2-h^2
+A\left(\ell^2-\frac{4M_N^2}{9}\right)
\nonumber\\
&+M^2(u+y)+v\Lambda_\Gamma^2.
\end{align}
Since $A\geq0$ and
$h^2\leq(1-2\beta)^2M_N^2/9$, this implies
\begin{align}
\operatorname{Re}D
\geq{}&
(u+y)\left(M^2-\frac{M_N^2}{9}\right)
\nonumber\\
&+v\left(\Lambda_\Gamma^2
-\frac{z^2M_N^2}{9}\right).
\end{align}
The right-hand side is strictly positive if
\begin{equation}
M>\frac{M_N}{3},
\quad
\Lambda_\Gamma>\frac{M_N}{3}.
\end{equation}
These sufficient conditions hold for the scalar-sector parameters used in this work. Therefore, the denominator does not vanish anywhere in the strip swept out by the deformation. For convergent loop integrals, the contour can be closed by connecting the two horizontal lines at infinity, provided that the connecting contributions vanish. Cauchy's theorem then equates the displaced and real-axis integrals, with no additional residue
terms. For moments requiring ultraviolet regularization, this argument must be implemented with a prescription that preserves translations and rotationally invariant tensor reduction \cite{Eichmann:2019dts,Bhagwat:2002tx}. After the contour deformation, the numerator reads
\begin{equation}
[(k'+\beta K)\cdot n]^m
=\sum_{j=0}^{m}\binom{m}{j}
(k'\cdot n)^j(\beta K\cdot n)^{m-j}.
\end{equation}
For a scalar weight $F(k'^2)$, odd powers of $k'\cdot n$ vanish by symmetric integration. The quadratic term satisfies
\begin{align}
\int d^4k'\,(k'\cdot n)^2F(k'^2)
&=\frac{n^2}{4}
\int d^4k'\,k'^2F(k'^2)
\nonumber\\
&=0.
\end{align}
Higher even-rank tensor integrals reduce to sums of products of Kronecker deltas. Contracting these with the light-like vector $n$ produces powers
of $n^2$, so all positive even powers vanish as well. Consequently,
\begin{align}
&\int d^4k'\,
[(k'+\beta K)\cdot n]^m F(k'^2)
\nonumber\\
&\qquad =
\beta^m(K\cdot n)^m
\int d^4k'\,F(k'^2),
\end{align}
where the identities are understood within the same convergence or regularization prescription.

\section{Eigensystem of the leading-order evolution kernel}
\label{app:evolution-kernel}

For completeness, in this appendix we provide the change-of-basis matrices, anomalous dimensions, polynomial norms, and projected moments used to evolve the normalized proton PDA from $\mu_0=0.51\,\mathrm{GeV}$ to $\mu=2\,\mathrm{GeV}$.  These quantities correspond to the polynomial convention adopted in Eq.~\eqref{eq:evol-basis}, in which $x_1$ denotes the momentum fraction of the bystander quark and $(x_2,x_3)$ are associated with the correlated pair.  In particular, the Gegenbauer argument is chosen as $(x_3-x_2)/(x_2+x_3)$.  Changing its sign changes the sign of the odd-$n$ basis elements and must be accompanied by the corresponding change in the projected moments and mixing matrices.

At every fixed total degree $N$, we collect the projected moments in
\begin{equation}
  \boldsymbol{B}_N
  =
  (B_{N0},B_{N1},\ldots,B_{NN})^{\mathrm T}.
\end{equation}
The multiplicatively renormalizable eigenmoments are defined by
\begin{equation}
  \boldsymbol{E}_N
  =
  \boldsymbol{V}_N\boldsymbol{B}_N,
  \qquad
  \boldsymbol{B}_N
  =
  \boldsymbol{V}_N^{-1}\boldsymbol{E}_N.
  \label{eq:app-change-basis}
\end{equation}
Their scale dependence is
\begin{equation}
  E_{Nk}(\mu)
  =
  E_{Nk}(\mu_0)
  \left[
    \frac{\ln(\mu^2/\Lambda_{\mathrm{QCD}}^2)}
         {\ln(\mu_0^2/\Lambda_{\mathrm{QCD}}^2)}
  \right]^{-[\gamma_{Nk}-\gamma_{00}]/\beta_0},
  \label{eq:app-evolution}
\end{equation}
where $\gamma_{00}=2/3$, $\beta_0=25/3$, and $\Lambda_{\mathrm{QCD}}=0.234\,\mathrm{GeV}$. The numerical entries below are quoted with the precision used in the evolution code; digits beyond the displayed precision should not be regarded as exact.

\subsection{Change-of-basis matrices and anomalous dimensions}

For $N=1$ and $N=2$, the matrices and anomalous dimensions are
\begin{align}
  \boldsymbol{V}_1
  &=
  \begin{pmatrix}
    0.6&1\\
    -1.8&1
  \end{pmatrix},
  &
  \boldsymbol{\gamma}_1
  &=
  \begin{pmatrix}
    10/3\\
    26/9
  \end{pmatrix},
  \label{eq:app-N1}
  \\
  \boldsymbol{V}_2
  &=
  \begin{pmatrix}
    -0.285714&1.42857&1\\
    -1&-0.714286&1\\
    1.5&-0.357143&1
  \end{pmatrix},
  &
  \boldsymbol{\gamma}_2
  &=
  \begin{pmatrix}
    16/3\\
    46/9\\
    38/9
  \end{pmatrix}.
  \label{eq:app-N2}
\end{align}
For the remaining degrees retained in the reconstruction, we find
\begin{equation}
\resizebox{0.96\linewidth}{!}{$
  \boldsymbol{V}_3
  =
  \begin{pmatrix}
    -0.0697889&-1.32104&-0.0271401&1\\
    -2.47481&0.555556&5.146&1\\
    0.808143&0.555556&0.039185&1\\
    -2.1228&0.674948&-0.825535&1
  \end{pmatrix},
  \quad
  \boldsymbol{\gamma}_3
  =
  \begin{pmatrix}
    6.98099\\
    6.93605\\
    5.84173\\
    5.44123
  \end{pmatrix}.
$}
\label{eq:app-N3}
\end{equation}
\begin{equation}
\resizebox{\linewidth}{!}{$
  \boldsymbol{V}_4
  =
  \begin{pmatrix}
    0.958959&3.1619&-2.54947&-3.86609&1\\
    0.495186&-0.792887&-1.49519&0.704353&1\\
    -0.24909&1.08124&0.239248&0.754877&1\\
    -1.21667&-0.521163&0.216673&-0.518403&1\\
    1.78741&-0.359872&0.73416&-0.35438&1
  \end{pmatrix},
  \quad
  \boldsymbol{\gamma}_4
  =
  \begin{pmatrix}
    8.42303\\
    8.41952\\
    7.25862\\
    6.95825\\
    6.31835
  \end{pmatrix}.
$}
\label{eq:app-N4}
\end{equation}
\begin{equation}
\resizebox{\linewidth}{!}{$
  \boldsymbol{V}_5
  =
  \begin{pmatrix}
    -0.276405&1.47975&1.15158&-2.31931&-1.20198&1\\
    0.737542&0.936982&-2.82881&-1.72336&2.28933&1\\
    -0.0808179&-1.1696&0.0256017&-0.272762&-0.058361&1\\
    -1.93144&0.530756&1.95316&0.290421&2.66737&1\\
    0.980537&0.459797&0.081271&0.485559&-0.0790375&1\\
    -2.33159&0.579202&-0.790157&0.574115&-0.648857&1
  \end{pmatrix},
  \quad
  \boldsymbol{\gamma}_5
  =
  \begin{pmatrix}
    9.67631\\
    9.67479\\
    8.47678\\
    8.40790\\
    7.51579\\
    7.16272
  \end{pmatrix}.
$}
\label{eq:app-N5}
\end{equation}

\subsection{Polynomial norms}
The squared norms defined in Eq.~\eqref{eq:evol-norm} are
\begin{subequations}
\begin{align}
  \boldsymbol{\mathcal N}_1
  &=
  \left(\frac{200}{7},\frac{216}{7}\right),
  \\
  \boldsymbol{\mathcal N}_2
  &=
  \left(40,56,\frac{320}{7}\right),
  \\
  \boldsymbol{\mathcal N}_3
  &=
  \left(
    \frac{560}{11},
    \frac{864}{11},
    \frac{6480}{77},
    \frac{2000}{33}
  \right),
  \\
  \boldsymbol{\mathcal N}_4
  &=
  \left(
    \frac{800}{13},
    \frac{1296}{13},
    \frac{10800}{91},
    \frac{4400}{39},
    \frac{10800}{143}
  \right),
  \\
  \boldsymbol{\mathcal N}_5
  &=
  \left(
    72,
    120,
    \frac{1056}{7},
    160,
    \frac{1560}{11},
    \frac{1176}{13}
  \right).
  \label{eq:app-norms}
\end{align}
\end{subequations}

\subsection{Projected moments before and after evolution}

For the proton PDA obtained with the $65\%$ scalar-diquark and $35\%$ axial-vector-diquark combination, the projected moments at the two scales are
\begin{subequations}
\label{eq:app-projected-moments}

\begin{align}
\boldsymbol{B}_1(\mu_0)
&=
\begin{pmatrix}
2.11034\\
1.17464
\end{pmatrix},
&
\boldsymbol{B}_1(2\,\mathrm{GeV})
&=
\begin{pmatrix}
1.56993\\
0.823079
\end{pmatrix},
\\[1ex]
\boldsymbol{B}_2(\mu_0)
&=
\begin{pmatrix}
1.50197\\
0.565861\\
1.10637
\end{pmatrix},
&
\boldsymbol{B}_2(2\,\mathrm{GeV})
&=
\begin{pmatrix}
0.974316\\
0.361808\\
0.720883
\end{pmatrix}.
\end{align}

\begin{align}
\boldsymbol{B}_3(\mu_0)
&=
\begin{pmatrix}
0.609137\\
0.431748\\
0.150788\\
0.617797
\end{pmatrix},
&
\boldsymbol{B}_3(2\,\mathrm{GeV})
&=
\begin{pmatrix}
0.322125\\
0.231096\\
0.0635048\\
0.330232
\end{pmatrix},
\\[1ex]
\boldsymbol{B}_4(\mu_0)
&=
\begin{pmatrix}
0.464972\\
0.285629\\
0.698691\\
0.638891\\
0.550222
\end{pmatrix},
&
\boldsymbol{B}_4(2\,\mathrm{GeV})
&=
\begin{pmatrix}
0.242971\\
0.151296\\
0.336791\\
0.292150\\
0.279012
\end{pmatrix}.
\end{align}

\begin{align}
\boldsymbol{B}_5(\mu_0)
&=
\begin{pmatrix}
0.146184\\
0.132239\\
0.303739\\
0.448499\\
0.508630\\
0.336689
\end{pmatrix},
&
\boldsymbol{B}_5(2\,\mathrm{GeV})
&=
\begin{pmatrix}
0.0550717\\
0.0686391\\
0.128869\\
0.187653\\
0.205653\\
0.154216
\end{pmatrix}.
\end{align}

\end{subequations}
The reduction of the nonconstant projected moments reflects the suppression of non-asymptotic structures as the renormalization scale increases. Using Eqs.~\eqref{eq:app-change-basis}--\eqref{eq:app-projected-moments} in the reconstruction formula of Eq.~\eqref{eq:evolved-reconstruction} reproduces the degree-five PDA shown in Fig.~\ref{fig:pda-evolution}.


\bibliography{print_ProtonPDA}

\end{document}